# Electron Affinity, Electronegativity and Electrophilicity of Anions

**P. K. Chattaraj* and S. Duley**

Department of Chemistry and Center for Theoretical Studies, Indian Institute of Technology, Kharagpur, 721302, India.

*To whom correspondence should be addressed: E-mail- pkc@chem.iitkgp.ernet.in

**Abstract**

Electron affinity, electronegativity and electrophilicity of several neutral atoms and their positive and negative ions are calculated at various levels of theory using different basis sets in the gas phase as well as in the presence of solvent and counterions. Electron affinity and electronegativity of all the anions and dianions are negative in gas phase and accordingly the electrophilicity is unexpectedly large vis – a - vis its quadratic definition. Many of these trends get altered in case the effects of solvent and counterions are taken into account.

## 1. Introduction

Electron affinity, electronegativity and electrophilicity are three related chemical concepts[1-3]. The difference in energy of a neutral atom and its anion in gas phase is the electron affinity (A) which may be equated with the electron-gain enthalpy with a minus sign, at T=0K. Therefore, the electron affinity of an N-electron system is given by,

$$A = E(N) - E(N+1) \qquad (1)$$

Electronegativity of an atom in a molecule is the power with which it attracts electrons to itself [4]. In order to provide an absolute definition of electronegativity ($\chi$) of an isolated species like atom, ion, molecule or solid, Mulliken[5] defined it as

$$\chi = \frac{I + A}{2} \qquad (2)$$

where I is the ionization potential given by

$$I = E(N-1) - E(N) \ . \qquad (3)$$

It implies that a system with larger I and A values would prefer to accept an electron rather than loosing it. The energy of a system may be approximately expressed as a quadratic function of the charge and it attains its minimum value for majority of the atoms / ions around the mononegative charge[1-3]. This fact is to be verified in the present work. Considering the slope of this parabola to be the electronegativity ($\chi$) it is easy to show that $\chi$ of a system also changes with the charges on it[6] which may become negative for the negatively charged species as the hardness ($\eta$, see below) is always positive due to the convexity of the E vs N plot[7-10]. Electrophilicity of a system is the measure of its reactivity towards attracting electrons from a nucleophile so that they form a bond. Inspired by the work of Maynard et al[11], a definition of an electrophilicity index (ω) is proposed by Parr et al[12-14] as,

$$\omega = \frac{\mu^2}{2\eta} = \frac{\chi^2}{2\eta} \tag{4}$$

where μ is the chemical potential[15] (the negative of the electronegativity ) and η is the chemical hardness[16]. The energy lowering due to the maximum amount of electron flow, which may be more or less than one as opposed to exactly one in the definition of A (eq 1), to a system from a free electron gas at 0K with μ = 0. Although Maynard et al[7] provided the empirical definition based on kinetic data the same definition (eq 4) is obtained by Parr et al[12-14] from an energy viewpoint.

Most of the neutral atoms possess positive A values and the halogen atoms have typically high A values as they attain noble gas (with very small or negative A values) configuration after accepting an electron. Anions possess negative A values since in these systems electron–electron repulsion outweighs the electron-nuclear attraction. Formation of stable metal oxides or sulphides is generally explained in terms of the role played by the lattice energy and solvation energy[1,2]. Pearson[17-19] has shown that the electronegativity values are more or less same in the gas and the solution phases. However, the corresponding hardness values decrease on solvation.

In the present work we calculate energy, electron affinity, ionization potential, electronegativity, hardness and electrophilicity of some selected atoms and their cations, dications, anions and dianions to analyze the electron accepting characteristics of those systems. Section 2 provides the numerical details while the results and discussion are presented in section 3. Finally section 4 contains some concluding remarks.

## 2. Numerical Details

All the calculations are done at the HF/6-311+G(d), B3LYP/6-311+G(d) and MP2/6-311+G(d) levels of theory. The I and A values are calculated using eqs (3) and (1) respectively, $\chi$ using eq (2), $\eta$ as[20] (I-A) , and $\omega$ using eq (4). We also use Koopmans' theorem to approximate I and A in terms of the appropriate frontier orbital energies. Calculations are also performed in the solution phase[21], in the presence of counter ions as well as with different basis sets. Electrodonating ($\omega^-$) and electroaccepting ($\omega^+$) powers[22] are also calculated in terms of $\mu^- = - I$, $\mu^+ = - A$ and $\eta^+ = \eta^- = \eta = (\mu^+ - \mu^-)$.

## 3. Results and Discussion

Tables 1 and 2 present the calculation of the energy, ionization potential, electron affinity, electronegativity, chemical hardness and electrophilicity of selected atoms/ions in the gas phase and in aqueous phase respectively. The calculations are done by using the Koopmans' theorem through the energies of the associated frontier orbitals, at B3LYP/6-311+G(d) level of theory and the tables 3-5 present the energy, ionization potential, electron affinity, electronegativity, chemical hardness and electrophilicity of selected atoms/ions in the gas phase calculated from the $\Delta$SCF using HF, MP2 and B3LYP levels of theory respectively. Koopmans' theorem can reproduce the expected trends in most cases but for Li and F. In case of Li, I value is overestimated while it is underestimated in case of F. Both cations and dications are highly electronegative and electrophilic, as expected. For anions and dianions both I and A and hence $\chi$ values are negative. It implies that they will not like to accept electrons. It may be noted that their $\omega$ values are very high which is counter-intuitive and definitely a drawback of the quadratic appearance of $\chi$ in eq 4. Tables 6-8 report all these quantities in the aqueous phase. For the neutral atoms and their cations and dications the numerical values differ but the trends remain more or less the same as that obtained in the gas phase. However, for the anions and the dianions situation changes drastically. Calculations do not converge for $N^{2-}$ in aqueous phase for the cases of HF and MP2 levels of theory. The $\chi$ values become positive in several systems and the $\omega$ values are no longer large. Tables 3-8 also present the two parabola model[14] results for electrodonating and electroaccepting processes in the gas and aqueous phases respectively. It may be noted that[15] while larger $\omega^+$ implies better accepting power smaller $\omega^-$ implies better donating power. In general $\omega^+$ follows the trend (for an element X): $X<X^+<X^{2+}$ (also $X<X^-<X^{2-}$) and $\omega^-$ follows the trend : $X^-<X<X^{2-}$ (also $X<X^+<X^{2+}$). The anomaly in these trends may be rationalized in the cases with positive $\mu^{\pm}$ values (negative $\chi^{\pm}$ values) and the quadratic appearance in the formula[22]: $\omega^{\pm} = \frac{(\mu^{\pm})^2}{2\eta}$

In order to check the inadequacy of the Koopmans' approximation we calculate the I and A values using eqs 1 and 3. Most of the important calculated quantities are provided in Tables 3-8. We use HF/6-311+G(d), MP2/6-311+G(d) and B3LYP/6-311+G(d) levels and also other basis sets like 6-31+G(d) and 6-311++G(d) for both in gas phase and also in aqueous phase calculations. A good agreement is found in gas phase calculation for atoms and their corresponding positive ions with experimental values. The use of B3LYP/6-311+G(d) level of theory to calculate ionization potential and electron affinity for the atoms and their cations and dications provides the best correlation with the experimental values. Table 9 presents the comparison between the calculated and the experimental values wherever available. For any system (except Be and Ne), energy becomes a minimum for the mononegative ion (Fig.1). The A value of the N – electron system is same as the I value of the corresponding (N+1) - electron system (vide eqs 1 and 3) which is not obeyed when Koopmans' approximation is used. It may be noted that this approximation should strictly be applied within the HF theory. As electrons are taken out the I, A, $\chi$ ,and $\omega$ values increase implying that it is difficult to eject electrons further and the system would rather prefer to accept electrons. For the anions the A and $\chi$ values are negative implying that they do not prefer to accept electrons any more as the electron – electron repulsion becomes stronger than the electron – nuclear attraction. However, large $\omega$ values for the dianions are surely counter-intuitive and are arising out of the quadratic appearance of $\chi$ in the expression for $\omega$ (eq 4). This problem persists in the ($\omega^+$, $\omega^-$) values apart from their problems mentioned above. The $\omega^+$ values of dianions are very large and are larger than the related $\omega^-$ values.

The dianions are unstable in gaseous phase and that can be stabilized by considering the presence of suitable counterions[23-26]. We calculate the ionization potential and the electron affinity for $M^{2-}(Z^+)_2$ : M=Li – Ne, molecules where Z contains one unit of point positive charge. Tables 10 - 12 present the values of the ionization potential, electron affinity, electronegativity, hardness, electrophilicity and also that of ($\mu^+$, $\mu^-$ , $\omega^+$, $\omega^-$) for the electroaccepting and electrodonating processes of the dianions in the presence of counterions, calculated at the HF, MP2 and B3LYP levels of theory respectively with the 6-311+G(d) basis set. Due to the presence of positive counterions the otherwise negative values of the ionization potential and the electron affinity of all the dianions become positive.

## 4. Concluding Remarks

It has been demonstrated through the calculation of ionization potential and electron affinity of several neutral atoms and their cations, dications, anions and dianions at the gas and solution phases at various levels of theory using different basis sets that the cations prefer to accept electrons while anions prefer to donate electrons. Mononegative ion is the most stable species of any element (except Be and Ne). Calculated values of electron affinity, electronegativity and electrophilicity of dianions often provide some counter-intuitive trends. Presence of counterions and/or solvent often remedies these problems

**Acknowledgments:** We are grateful to Professor Ralph G. Pearson and Mr. S. Giri for helpful discussions and the Indo - EU project (HYPOMAP) for financial assistance.

**Table 1**: Ionization potential(I) , Electron affinity(A), Electronegativity(χ), Chemical hardness(η) , Electrophilicity(ω) and the values of ($\mu^+$, $\mu^-$ ,$\omega^+$ ,$\omega^-$ ) for Electroaccepting and Electrodonating processes of atoms and ions using Koopmans' theorem al B3LYP/6-11+G(d) level of theory

(a)

| Atoms | I (eV) | A (eV) | χ (eV) | η (eV) | ω (eV) | $\mu^+$ (eV) | $\mu^-$ (eV) | η (eV) | $\omega^+$ (eV) | $\omega^-$ (eV) |
|---|---|---|---|---|---|---|---|---|---|---|
| Li | 29.308 | 1.172 | 15.240 | 28.136 | 4.128 | -1.172 | -29.308 | 28.136 | 0.024 | 15.265 |
| Be | 6.318 | 1.430 | 3.874 | 4.889 | 1.535 | -1.430 | -6.318 | 4.888 | 0.209 | 4.083 |
| B | 7.386 | 2.571 | 4.978 | 4.814 | 2.574 | -2.571 | -7.386 | 4.815 | 0.687 | 5.665 |
| C | 6.017 | 4.324 | 5.171 | 1.693 | 7.897 | -4.324 | -6.017 | 1.693 | 5.523 | 10.693 |
| N | 7.886 | 5.944 | 6.915 | 1.943 | 12.307 | -5.944 | -7.886 | 1.942 | 9.092 | 16.007 |
| O | 10.385 | 8.192 | 9.288 | 2.193 | 19.670 | -8.192 | -10.385 | 2.193 | 15.300 | 24.588 |
| F | 12.727 | 2.302 | 7.514 | 10.425 | 2.708 | -2.302 | -12.727 | 10.425 | 0.254 | 7.768 |
| Ne | 15.693 | -3.964 | 5.864 | 19.656 | 0.875 | 3.964 | -15.693 | 19.657 | 0.400 | 6.264 |

(b)

| Ions | I (eV) | A (eV) | χ (eV) | η (eV) | ω (eV) | $\mu^+$ (eV) | $\mu^-$ (eV) | η (eV) | $\omega^+$ (eV) | $\omega^-$ (eV) |
|---|---|---|---|---|---|---|---|---|---|---|
| $Li^+$ | 63.914 | 6.942 | 35.428 | 56.971 | 11.015 | -6.942 | -63.914 | 56.972 | 0.423 | 35.851 |
| $Be^+$ | 68.764 | 10.518 | 39.641 | 58.246 | 13.489 | -10.518 | -68.764 | 58.246 | 0.950 | 40.591 |
| $B^+$ | 20.576 | 12.378 | 16.477 | 8.197 | 16.560 | -12.378 | -20.576 | 8.198 | 9.346 | 25.823 |
| $C^+$ | 68.764 | 10.518 | 39.641 | 58.246 | 13.489 | -10.518 | -68.764 | 58.246 | 0.950 | 40.591 |
| $N^+$ | 21.619 | 19.357 | 20.488 | 2.262 | 92.770 | -19.357 | -21.619 | 2.262 | 82.809 | 103.297 |
| $O^+$ | 25.517 | 23.039 | 24.278 | 2.478 | 118.914 | -23.039 | -25.517 | 2.478 | 107.085 | 131.363 |
| $F^+$ | 30.017 | 27.331 | 28.674 | 2.686 | 153.049 | -27.331 | -30.017 | 2.686 | 139.048 | 167.722 |
| $Ne^+$ | 34.383 | 16.930 | 25.656 | 17.453 | 18.857 | -16.930 | -34.383 | 17.453 | 8.211 | 33.867 |

(c)

| Ions | I (eV) | A (eV) | χ (eV) | η (eV) | ω (eV) | $\mu^+$ (eV) | $\mu^-$ (eV) | η (eV) | $\omega^+$ (eV) | $\omega^-$ (eV) |
|---|---|---|---|---|---|---|---|---|---|---|
| $Li^{2+}$ | 60.721 | 45.921 | 53.321 | 14.800 | 96.048 | -45.921 | -60.721 | 14.800 | 71.238 | 124.559 |
| $Be^{2+}$ | 137.136 | 21.395 | 79.265 | 115.741 | 27.143 | -21.395 | -137.136 | 115.741 | 1.977 | 81.243 |
| $B^{2+}$ | 125.330 | 26.798 | 76.064 | 98.532 | 29.360 | -26.798 | -125.330 | 98.532 | 3.644 | 79.708 |
| $C^{2+}$ | 41.782 | 30.441 | 36.112 | 11.341 | 57.492 | -30.441 | -41.782 | 11.341 | 40.854 | 76.966 |
| $N^{2+}$ | 45.182 | 35.684 | 40.433 | 9.498 | 86.063 | -35.684 | -45.182 | 9.498 | 67.034 | 107.467 |
| $O^{2+}$ | 44.333 | 41.511 | 42.922 | 2.821 | 326.499 | -41.511 | -44.333 | 2.821 | 305.391 | 348.313 |
| $F^{2+}$ | 50.268 | 47.268 | 48.768 | 2.999 | 396.487 | -47.268 | -50.268 | 2.999 | 372.478 | 421.246 |
| $Ne^{2+}$ | 56.792 | 53.623 | 55.207 | 3.169 | 480.918 | -53.623 | -56.792 | 3.169 | 453.710 | 508.917 |

(d)

| Ions | I (eV) | A (eV) | χ (eV) | η (eV) | ω (eV) | μ⁺ (eV) | μ⁻ (eV) | η (eV) | ω⁺ (eV) | ω⁻ (eV) |
|---|---|---|---|---|---|---|---|---|---|---|
| $Li^-$ | -0.580 | -1.744 | -1.162 | 1.165 | 0.580 | 1.744 | 0.580 | 1.164 | 1.306 | 0.144 |
| $Be^-$ | -0.245 | -2.538 | -1.392 | 2.293 | 0.422 | 2.538 | 0.245 | 2.293 | 1.405 | 0.013 |
| $B^-$ | -1.946 | -3.044 | -2.495 | 1.098 | 2.835 | 3.044 | 1.946 | 1.098 | 4.220 | 1.724 |
| $C^-$ | -2.029 | -3.402 | -2.716 | 1.373 | 2.686 | 3.402 | 2.029 | 1.373 | 4.216 | 1.500 |
| $N^-$ | -1.609 | -3.283 | -2.446 | 1.674 | 1.788 | 3.283 | 1.609 | 1.674 | 3.220 | 0.774 |
| $O^-$ | -1.296 | -7.078 | -4.187 | 5.782 | 1.516 | 7.078 | 1.296 | 5.782 | 4.332 | 0.145 |
| $F^-$ | -0.353 | -11.062 | -5.708 | 10.709 | 1.521 | 11.062 | 0.353 | 10.709 | 5.713 | 0.006 |
| $Ne^-$ | -1.515 | -12.124 | -6.820 | 10.608 | 2.192 | 12.124 | 1.515 | 10.609 | 6.927 | 0.108 |

(e)

| Ions | I (eV) | A (eV) | χ (eV) | η (eV) | ω (eV) | μ⁺ (eV) | μ⁻ (eV) | η (eV) | ω⁺ (eV) | ω⁻ (eV) |
|---|---|---|---|---|---|---|---|---|---|---|
| $Li^{2-}$ | -2.398 | -3.101 | -2.750 | 0.703 | 5.376 | 3.101 | 2.398 | 0.703 | 6.839 | 4.089 |
| $Be^{2-}$ | -0.835 | -1.383 | -1.109 | 0.547 | 1.123 | 1.383 | 0.835 | 0.547 | 1.746 | 0.637 |
| $B^{2-}$ | 6.802 | 5.754 | 6.278 | 1.048 | 18.801 | -5.754 | -6.802 | 1.048 | 15.793 | 22.071 |
| $C^{2-}$ | -6.972 | -8.129 | -7.551 | 1.158 | 24.626 | 8.129 | 6.972 | 1.158 | 28.546 | 20.995 |
| $N^{2-}$ | -8.402 | -11.733 | -10.067 | 3.331 | 15.213 | 11.733 | 8.402 | 3.331 | 20.663 | 10.596 |
| $O^{2-}$ | -9.289 | -15.326 | -12.307 | 6.037 | 12.546 | 15.326 | 9.289 | 6.037 | 19.455 | 7.147 |
| $F^{2-}$ | -11.781 | -18.026 | -14.903 | 6.245 | 17.782 | 18.026 | 11.781 | 6.245 | 26.014 | 11.111 |
| $Ne^{2-}$ | -17.234 | -19.898 | -18.566 | 2.665 | 64.681 | 19.898 | 17.234 | 2.665 | 74.297 | 55.731 |

**Table 2**: Ionization potential(I) , Electron affinity(A), Electronegativity(χ), Chemical hardness(η) , Electrophilicity(ω) and the values of ($\mu^+$, $\mu^-$ ,$\omega^+$ ,$\omega^-$ ) for Electroaccepting and Electrodonating processes of atoms and ions using Koopmans' theorem in aqueous solution B3LYP/6-11+G(d) level of theory

(a)

| **Atoms** | **I (eV)** | **A (eV)** | **χ (eV)** | **η (eV)** | **ω (eV)** | **$\mu^+$ (eV)** | **$\mu^-$ (eV)** | **η (eV)** | **$\omega^+$ (eV)** | **$\omega^-$ (eV)** |
|---|---|---|---|---|---|---|---|---|---|---|
| Li | 29.310 | 1.173 | 15.241 | 28.137 | 4.129 | -1.173 | -29.310 | 28.137 | 0.024 | 15.266 |
| Be | 6.319 | 1.430 | 3.874 | 4.889 | 1.535 | -1.430 | -6.319 | 4.889 | 0.209 | 4.083 |
| B | 7.458 | 2.519 | 4.988 | 4.939 | 2.519 | -2.519 | -7.458 | 4.939 | 0.642 | 5.630 |
| C | 6.244 | 4.307 | 5.275 | 1.937 | 7.183 | -4.307 | -6.244 | 1.937 | 4.787 | 10.063 |
| N | 7.981 | 5.916 | 6.948 | 2.065 | 11.687 | -5.916 | -7.981 | 2.065 | 8.471 | 15.420 |
| O | 10.454 | 8.153 | 9.303 | 2.300 | 18.812 | -8.153 | -10.454 | 2.301 | 14.448 | 23.751 |
| F | 12.746 | 2.291 | 7.519 | 10.455 | 2.703 | -2.291 | -12.746 | 10.455 | 0.251 | 7.770 |
| Ne | 15.693 | -3.963 | 5.865 | 19.656 | 0.875 | 3.963 | -15.693 | 19.656 | 0.400 | 6.264 |

(b)

| **Ions** | **I (eV)** | **A (eV)** | **χ (eV)** | **η (eV)** | **ω (eV)** | **$\mu^+$ (eV)** | **$\mu^-$ (eV)** | **η (eV)** | **$\omega^+$ (eV)** | **$\omega^-$ (eV)** |
|---|---|---|---|---|---|---|---|---|---|---|
| $Li^+$ | 55.273 | 0.150 | 27.711 | 55.123 | 6.965 | -0.150 | -55.273 | 55.123 | 0.000 | 27.711 |
| $Be^+$ | 60.951 | 2.865 | 31.908 | 58.086 | 8.764 | -2.865 | -60.951 | 58.086 | 0.071 | 31.979 |
| $B^+$ | 12.541 | 4.440 | 8.491 | 8.101 | 4.450 | -4.440 | -12.541 | 8.101 | 1.217 | 9.708 |
| $C^+$ | 14.959 | 7.737 | 11.348 | 7.222 | 8.916 | -7.737 | -14.959 | 7.222 | 4.145 | 15.493 |
| $N^+$ | 14.464 | 12.126 | 13.295 | 2.338 | 37.801 | -12.126 | -14.464 | 2.338 | 31.445 | 44.740 |
| $O^+$ | 16.156 | 13.526 | 14.841 | 2.630 | 41.881 | -13.526 | -16.156 | 2.630 | 34.789 | 49.631 |
| $F^+$ | 22.151 | 19.419 | 20.785 | 2.731 | 79.089 | -19.419 | -22.151 | 2.732 | 69.038 | 89.823 |
| $Ne^+$ | 25.296 | 8.128 | 16.712 | 17.167 | 8.134 | -8.128 | -25.296 | 17.168 | 1.924 | 18.636 |

(c)

| **Ions** | **I (eV)** | **A (eV)** | **χ (eV)** | **η (eV)** | **ω (eV)** | **$\mu^+$ (eV)** | **$\mu^-$ (eV)** | **η (eV)** | **$\omega^+$ (eV)** | **$\omega^-$ (eV)** |
|---|---|---|---|---|---|---|---|---|---|---|
| $Li^{2+}$ | 90.156 | 29.442 | 59.799 | 60.713 | 29.449 | -29.442 | -90.156 | 60.714 | 7.139 | 66.938 |
| $Be^{2+}$ | 120.797 | 5.259 | 63.028 | 115.538 | 17.192 | -5.259 | -120.797 | 115.538 | 0.120 | 63.1480 |
| $B^{2+}$ | 109.065 | 10.560 | 59.813 | 98.505 | 18.159 | -10.560 | -109.065 | 98.505 | 0.566 | 60.379 |
| $C^{2+}$ | 26.009 | 14.682 | 20.346 | 11.327 | 18.272 | -14.682 | -26.009 | 11.327 | 9.515 | 29.861 |
| $N^{2+}$ | 30.703 | 21.190 | 25.947 | 9.513 | 35.385 | -21.190 | -30.703 | 9.513 | 23.601 | 49.547 |
| $O^{2+}$ | 20.757 | 17.662 | 19.210 | 3.095 | 59.609 | -17.662 | -20.757 | 3.095 | 50.391 | 69.601 |
| $F^{2+}$ | 34.493 | 31.462 | 32.978 | 3.031 | 179.372 | -31.462 | -34.493 | 3.031 | 163.262 | 196.24 |
| $Ne^{2+}$ | 38.591 | 35.380 | 36.985 | 3.211 | 212.990 | -35.380 | -38.591 | 3.211 | 194.899 | 231.884 |

(d)

| Ions | I (eV) | A (eV) | χ (eV) | η (eV) | ω (eV) | $\mu^+$ (eV) | $\mu^-$ (eV) | η (eV) | $\omega^+$ (eV) | $\omega^-$ (eV) |
|---|---|---|---|---|---|---|---|---|---|---|
| $Li^-$ | 4.417 | 1.435 | 2.926 | 2.982 | 1.436 | -1.435 | -4.417 | 2.982 | 0.345 | 3.272 |
| $Be^-$ | 5.010 | 1.850 | 3.430 | 3.160 | 1.862 | -1.850 | -5.010 | 3.160 | 0.542 | 3.972 |
| $B^-$ | 4.058 | 2.526 | 3.292 | 1.532 | 3.536 | -2.526 | -4.058 | 1.532 | 2.082 | 5.374 |
| $C^-$ | 4.549 | 2.841 | 3.695 | 1.708 | 3.998 | -2.841 | -4.549 | 1.708 | 2.364 | 6.059 |
| $N^-$ | 5.738 | 3.797 | 4.767 | 1.940 | 5.856 | -3.797 | -5.738 | 1.941 | 3.715 | 8.482 |
| $O^-$ | 7.423 | 0.769 | 4.096 | 6.654 | 1.261 | -0.769 | -7.423 | 6.654 | 0.044 | 4.141 |
| $F^-$ | 8.736 | -3.010 | 2.863 | 11.746 | 0.349 | 3.010 | -8.736 | 11.746 | 0.386 | 3.248 |
| $Ne^-$ | 7.144 | -3.998 | 1.573 | 11.143 | 0.111 | 3.998 | -7.144 | 11.143 | 0.717 | 2.290 |

(e)

| Ions | I (eV) | A (eV) | χ (eV) | η (eV) | ω (eV) | $\mu^+$ (eV) | $\mu^-$ (eV) | η (eV) | $\omega^+$ (eV) | $\omega^-$ (eV) |
|---|---|---|---|---|---|---|---|---|---|---|
| $Li^{2-}$ | 4.379 | 2.198 | 3.289 | 2.181 | 2.480 | -2.198 | -4.379 | 2.181 | 1.108 | 4.397 |
| $Be^{2-}$ | 3.686 | 2.546 | 3.116 | 1.141 | 4.256 | -2.546 | -3.686 | 1.140 | 2.840 | 5.956 |
| $B^{2-}$ | 4.554 | 2.041 | 3.298 | 2.513 | 2.164 | -2.041 | -4.554 | 2.513 | 0.829 | 4.127 |
| $C^{2-}$ | 4.425 | 2.866 | 3.645 | 1.559 | 4.261 | -2.866 | -4.425 | 1.559 | 2.634 | 6.279 |
| $N^{2-}$ | 5.702 | 1.023 | 3.362 | 4.679 | 1.208 | -1.023 | -5.702 | 4.679 | 0.112 | 3.474 |
| $O^{2-}$ | 8.726 | -1.589 | 3.569 | 10.315 | 0.617 | 1.589 | -8.726 | 10.315 | 0.122 | 3.691 |
| $F^{2-}$ | 7.263 | -2.191 | 2.536 | 9.455 | 0.340 | 2.191 | -7.263 | 9.454 | 0.254 | 2.789 |
| $Ne^{2-}$ | -0.530 | -4.215 | -2.372 | 3.685 | 0.764 | 4.215 | 0.530 | 3.685 | 2.411 | 0.038 |

**Table 3**: Ionization potential(I) , Electron affinity(E), Electronegativity(χ), Chemical hardness(η) , Electrophilicity(ω) and the values of ($\mu^+$, $\mu^-$ ,$\omega^+$ ,$\omega^-$ ) for Electroaccepting and Electrodonating processes of atoms and ions by using Hartree-Fock method

(a)

| Atoms | I (eV) | E (eV) | χ (eV) | η (eV) | ω (eV) | $\mu^+$ (eV) | $\mu^-$ (eV) | η (eV) | $\omega^+$ (eV) | $\omega^-$ (eV) | Basis sets |
|---|---|---|---|---|---|---|---|---|---|---|---|
| Li | 5.334 | -0.156 | 2.589 | 5.489 | 0.616 | 0.156 | -5.334 | 5.450 | 0.002 | 2.591 | 6-31+g(d) |
| | 5.338 | -0.124 | 2.601 | 5.462 | 0.622 | 0.124 | -5.338 | 5.462 | 0.001 | 2.609 | 6-311+g(d) |
| | 5.338 | -0.124 | 2.607 | 5.462 | 0.622 | 0.124 | -5.338 | 5.462 | 0.001 | 2.609 | 6-311++g(d) |
| Be | 8.006 | -0.708 | 3.649 | 8.713 | 0.764 | 0.708 | -8.006 | 8.714 | 0.029 | 3.678 | 6-31+g(d) |
| | 8.047 | -0.703 | 3.672 | 8.751 | 0.771 | 0.703 | -8.047 | 8.750 | 0.028 | 3.700 | 6-311+g(d) |
| | 8.047 | -0.703 | 3.672 | 8.751 | 0.771 | 0.703 | -8.047 | 8.750 | 0.028 | 3.700 | 6-311++g(d) |
| B | 7.947 | -1.542 | 3.203 | 9.490 | 0.540 | 1.542 | -7.947 | 9.489 | 0.125 | 3.328 | 6-31+g(d) |
| | 8.022 | -1.574 | 3.224 | 9.596 | 0.542 | 1.574 | -8.022 | 9.596 | 0.129 | 3.353 | 6-311+g(d) |
| | 8.022 | -1.574 | 3.224 | 9.596 | 0.542 | 1.574 | -8.022 | 9.596 | 0.129 | 3.353 | 6-311++g(d) |
| C | 8.357 | 0.700 | 4.528 | 7.657 | 1.339 | -0.700 | -8.357 | 7.657 | 0.032 | 4.561 | 6-31+g(d) |
| | 8.378 | 0.700 | 4.536 | 7.684 | 1.339 | -0.700 | -8.378 | 7.678 | 0.031 | 4.568 | 6-311+g(d) |
| | 8.022 | 0.700 | 4.536 | 7.684 | 1.339 | -0.700 | -8.022 | 7.322 | 0.031 | 4.568 | 6-311++g(d) |
| N | 13.622 | -0.716 | 6.452 | 14.337 | 1.452 | 0.716 | -13.622 | 14.338 | 0.018 | 6.471 | 6-31+g(d) |
| | 13.586 | -0.729 | 6.428 | 14.315 | 1.443 | 0.729 | -13.586 | 14.315 | 0.018 | 6.447 | 6-311+g(d) |
| | 13.585 | -0.729 | 6.428 | 14.315 | 1.443 | 0.729 | -13.585 | 14.315 | 0.018 | 6.447 | 6-311++g(d) |
| O | 13.617 | 2.895 | 8.256 | 10.722 | 3.179 | -2.895 | -13.617 | 10.722 | 0.391 | 8.647 | 6-31+g(d) |
| | 13.576 | 2.874 | 8.225 | 10.703 | 3.160 | -2.874 | -13.576 | 10.702 | 0.386 | 8.611 | 6-311+g(d) |
| | 13.576 | 2.874 | 8.225 | 10.703 | 3.160 | -2.874 | -13.576 | 10.702 | 0.386 | 8.611 | 6-311++g(d) |
| F | 20.106 | 1.277 | 10.691 | 18.828 | 3.035 | -1.277 | -20.106 | 18.829 | 0.043 | 10.735 | 6-31+g(d) |
| | 20.092 | 1.245 | 10.668 | 18.846 | 3.019 | -1.245 | -20.092 | 18.847 | 0.041 | 10.710 | 6-311+g(d) |
| | 20.092 | 1.245 | 10.668 | 18.846 | 3.019 | -1.245 | -20.092 | 18.847 | 0.041 | 10.709 | 6-311++g(d) |
| Ne | 19.891 | -7.852 | 6.019 | 27.743 | 0.653 | 7.852 | -19.891 | 27.743 | 1.111 | 7.131 | 6-31+g(d) |
| | 19.794 | -7.553 | 6.121 | 27.346 | 0.685 | 7.553 | -19.794 | 27.347 | 1.043 | 7.164 | 6-311+g(d) |
| | 19.794 | -7.553 | 6.121 | 27.346 | 0.685 | 7.553 | -19.794 | 27.347 | 1.043 | 7.164 | 6-311++g(d) |

(b)

| Ions | I (eV) | E (eV) | χ (eV) | η (eV) | ω (eV) | $\mu^+$ (eV) | $\mu^-$ (eV) | η (eV) | $\omega^+$ (eV) | $\omega^-$ (eV) | Basis sets |
|---|---|---|---|---|---|---|---|---|---|---|---|
| $Li^+$ | 75.644 | 5.334 | 40.489 | 70.310 | 11.658 | -5.334 | -75.644 | 70.310 | 0.202 | 40.692 | 6-31+g(d) |
| | 74.612 | 5.338 | 39.975 | 69.273 | 11.534 | -5.338 | -74.612 | 69.274 | 0.206 | 40.181 | 6-311+g(d) |
| | 74.612 | 5.338 | 39.975 | 69.273 | 11.534 | -5.338 | -74.612 | 69.274 | 0.206 | 40.181 | 6-311++g(d) |
| $Be^+$ | 18.115 | 8.005 | 13.061 | 10.110 | 8.436 | -8.005 | -18.115 | 10.110 | 3.170 | 16.230 | 6-31+g(d) |
| | 18.118 | 8.047 | 13.083 | 10.071 | 8.498 | -8.047 | -18.118 | 10.071 | 3.215 | 16.298 | 6-311+g(d) |
| | 18.118 | 8.047 | 13.083 | 10.071 | 8.498 | -8.047 | -18.118 | 10.071 | 3.215 | 16.298 | 6-311++g(d) |
| $B^+$ | 23.450 | 7.948 | 15.699 | 15.503 | 7.949 | -7.948 | -23.450 | 15.502 | 2.037 | 17.736 | 6-31+g(d) |
| | 23.442 | 8.022 | 15.732 | 15.420 | 8.025 | -8.022 | -23.442 | 15.420 | 2.086 | 17.818 | 6-311+g(d) |
| | 23.442 | 8.022 | 15.732 | 15.420 | 8.025 | -8.022 | -23.442 | 15.420 | 2.086 | 17.818 | 6-311++g(d) |
| $C^+$ | 24.137 | 8.357 | 16.247 | 15.780 | 8.364 | -8.357 | -24.137 | 15.780 | 2.213 | 18.460 | 6-31+g(d) |
| | 24.147 | 8.378 | 16.263 | 15.769 | 8.386 | -8.378 | -24.147 | 15.769 | 2.226 | 18.489 | 6-311+g(d) |
| | 24.147 | 8.378 | 16.263 | 15.769 | 8.386 | -8.378 | -24.147 | 15.769 | 2.226 | 18.489 | 6-311++g(d) |
| $N^+$ | 25.842 | 13.621 | 19.732 | 12.221 | 15.929 | -13.621 | -25.842 | 12.221 | 7.591 | 27.323 | 6-31+g(d) |
| | 25.783 | 13.586 | 19.684 | 12.197 | 15.884 | -13.586 | -25.783 | 12.197 | 7.566 | 27.251 | 6-311+g(d) |
| | 25.783 | 13.586 | 19.684 | 12.197 | 15.884 | -13.586 | -25.783 | 12.197 | 7.566 | 27.251 | 6-311++g(d) |
| $O^+$ | 33.925 | 13.617 | 23.771 | 20.308 | 13.912 | -13.617 | -33.925 | 20.308 | 4.565 | 28.337 | 6-31+g(d) |
| | 33.878 | 13.576 | 23.727 | 20.302 | 13.865 | -13.576 | -33.878 | 20.302 | 4.539 | 28.266 | 6-311+g(d) |
| | 33.878 | 13.576 | 23.727 | 20.302 | 13.865 | -13.576 | -33.878 | 20.302 | 4.539 | 28.266 | 6-311++g(d) |
| $F^+$ | 35.262 | 20.106 | 27.684 | 15.157 | 25.283 | -20.106 | -35.262 | 15.156 | 13.335 | 41.019 | 6-31+g(d) |
| | 35.186 | 20.092 | 27.638 | 15.094 | 25.304 | -20.092 | -35.186 | 15.094 | 13.372 | 41.010 | 6-311+g(d) |
| | 35.186 | 20.092 | 27.638 | 15.094 | 25.304 | -20.092 | -35.186 | 15.094 | 13.372 | 41.010 | 6-311++g(d) |
| $Ne^+$ | 44.591 | 19.891 | 32.241 | 24.700 | 21.042 | -19.891 | -44.591 | 24.700 | 8.009 | 40.250 | 6-31+g(d) |
| | 44.577 | 19.794 | 32.186 | 24.784 | 20.899 | -19.794 | -44.577 | 24.783 | 7.904 | 40.090 | 6-311+g(d) |
| | 44.577 | 19.794 | 32.186 | 24.784 | 20.899 | -19.794 | -44.577 | 24.783 | 7.904 | 40.090 | 6-311++g(d) |

(c)

| Ions | I (eV) | E (eV) | χ (eV) | η (eV) | ω (eV) | μ⁺ (eV) | μ⁻ (eV) | η (eV) | ω⁺ (eV) | ω⁻ (eV) | Basis sets |
|---|---|---|---|---|---|---|---|---|---|---|---|
| $Li^{2+}$ | 121.245 | 75.644 | 98.445 | 45.601 | 106.263 | -75.644 | -121.245 | 45.601 | 62.741 | 161.186 | 6-31+g(d) |
| | 122.286 | 74.612 | 98.449 | 47.674 | 101.650 | -74.612 | -122.286 | 47.674 | 58.385 | 156.834 | 6-311+g(d) |
| | 122.286 | 74.612 | 98.449 | 47.674 | 101.650 | -74.612 | -122.286 | 47.674 | 58.385 | 156.834 | 6-311++g(d) |
| $Be^{2+}$ | 153.250 | 18.115 | 85.682 | 135.134 | 27.164 | -18.115 | -153.250 | 135.135 | 1.214 | 86.897 | 6-31+g(d) |
| | 152.879 | 18.118 | 85.498 | 134.761 | 27.122 | -18.118 | -152.879 | 134.761 | 1.218 | 86.716 | 6-311+g(d) |
| | 152.879 | 18.118 | 85.498 | 134.761 | 27.122 | -18.118 | -152.879 | 134.761 | 1.218 | 86.716 | 6-311++g(d) |
| $B^{2+}$ | 37.781 | 23.451 | 30.616 | 14.330 | 32.705 | -23.451 | -37.781 | 14.330 | 19.188 | 49.804 | 6-31+g(d) |
| | 37.801 | 23.442 | 30.622 | 14.359 | 32.651 | -23.442 | -37.801 | 14.359 | 19.136 | 49.757 | 6-311+g(d) |
| | 37.801 | 23.442 | 30.622 | 14.359 | 32.651 | -23.442 | -37.801 | 14.359 | 19.136 | 49.757 | 6-311++g(d) |
| $C^{2+}$ | 45.762 | 24.137 | 34.950 | 21.624 | 28.243 | -24.137 | -45.762 | 21.625 | 13.471 | 48.421 | 6-31+g(d) |
| | 45.802 | 24.147 | 34.975 | 21.655 | 28.244 | -24.147 | -45.802 | 21.655 | 13.463 | 48.438 | 6-311+g(d) |
| | 45.802 | 24.147 | 34.975 | 21.655 | 28.244 | -24.147 | -45.802 | 21.655 | 13.463 | 49.438 | 6-311++g(d) |
| $N^{2+}$ | 47.271 | 25.842 | 36.557 | 21.429 | 31.182 | -25.842 | -47.271 | 21.429 | 15.582 | 52.139 | 6-31+g(d) |
| | 47.310 | 25.783 | 36.545 | 21.527 | 31.023 | -25.783 | -47.310 | 21.527 | 15.440 | 51.987 | 6-311+g(d) |
| | 47.310 | 25.783 | 36.545 | 21.527 | 31.023 | -25.783 | -47.310 | 21.527 | 15.440 | 51.987 | 6-311++g(d) |
| $O^{2+}$ | 50.378 | 33.925 | 42.152 | 16.453 | 53.996 | -33.925 | -50.378 | 16.453 | 34.977 | 77.129 | 6-31+g(d) |
| | 50.277 | 33.878 | 42.077 | 16.399 | 53.984 | -33.878 | -50.277 | 16.399 | 34.995 | 77.072 | 6-311+g(d) |
| | 50.277 | 33.878 | 42.077 | 16.398 | 53.984 | -33.878 | -50.277 | 16.399 | 34.995 | 77.073 | 6-311++g(d) |
| $F^{2+}$ | 61.241 | 35.262 | 48.252 | 25.978 | 44.811 | -35.262 | -61.241 | 25.979 | 23.932 | 72.184 | 6-31+g(d) |
| | 61.219 | 35.186 | 48.202 | 26.033 | 44.625 | -35.186 | -61.219 | 26.033 | 23.778 | 71.980 | 6-311+g(d) |
| | 61.219 | 35.186 | 48.202 | 26.033 | 44.625 | -35.186 | -61.219 | 26.033 | 23.778 | 71.980 | 6-311++g(d) |
| $Ne^{2+}$ | 63.935 | 44.591 | 54.263 | 19.344 | 76.109 | -44.591 | -63.935 | 19.344 | 51.396 | 105.658 | 6-31+g(d) |
| | 63.864 | 44.577 | 54.221 | 19.286 | 76.216 | -44.577 | -63.864 | 19.287 | 51.516 | 105.737 | 6-311+g(d) |
| | 63.864 | 44.577 | 54.221 | 19.286 | 76.216 | -44.577 | -63.864 | 19.287 | 51.516 | 105.737 | 6-311++g(d) |

(d)

| Ions | I (eV) | E (eV) | χ (eV) | η (eV) | ω (eV) | μ⁺ (eV) | μ⁻ (eV) | η (eV) | ω⁺ (eV) | ω⁻ (eV) | Basis sets |
|---|---|---|---|---|---|---|---|---|---|---|---|
| $Li^{-}$ | -0.156 | -2.130 | -1.143 | 1.974 | 0.331 | 2.130 | 0.156 | 1.974 | 1.149 | 0.006 | 6-31+g(d) |
| | -0.124 | -2.383 | -1.253 | 2.259 | 0.347 | 2.383 | 0.124 | 2.259 | 1.257 | 0.003 | 6-311+g(d) |
| | -0.124 | -2.383 | -1.253 | 2.259 | 0.347 | 2.383 | 0.124 | 2.259 | 1.257 | 0.003 | 6-311++g(d) |
| $Be^{-}$ | -0.708 | -4.541 | -2.624 | 3.883 | 0.898 | 4.541 | 0.708 | 3.883 | 2.690 | 0.065 | 6-31+g(d) |
| | -0.703 | -4.412 | -2.557 | 3.708 | 0.882 | 4.412 | 0.703 | 3.709 | 2.624 | 0.067 | 6-311+g(d) |
| | -0.703 | -4.412 | -2.557 | 3.708 | 0.882 | 4.412 | 0.703 | 3.709 | 2.624 | 0.067 | 6-311++g(d) |
| $B^{-}$ | -1.542 | -4.911 | -3.226 | 3.369 | 1.545 | 4.911 | 1.542 | 3.369 | 3.579 | 0.353 | 6-31+g(d) |
| | -1.574 | -4.902 | -3.238 | 3.328 | 1.575 | 4.902 | 1.574 | 3.328 | 3.610 | 0.372 | 6-311+g(d) |
| | -1.574 | -4.902 | -3.238 | 3.328 | 1.575 | 4.902 | 1.574 | 3.328 | 3.610 | 0.372 | 6-311++g(d) |
| $C^{-}$ | 0.700 | -7.334 | -3.317 | 8.034 | 0.685 | 7.334 | -0.700 | 8.034 | 3.347 | 0.030 | 6-31+g(d) |
| | 0.700 | -7.410 | -3.358 | 8.104 | 0.696 | 7.410 | -0.700 | 8.104 | 3.387 | 0.030 | 6-311+g(d) |
| | 0.700 | -7.410 | -3.358 | 8.104 | 0.696 | 7.410 | -0.700 | 8.104 | 3.387 | 0.030 | 6-311++g(d) |
| $N^{-}$ | -0.716 | -6.798 | -3.755 | 6.078 | 1.160 | 6.798 | 0.716 | 6.082 | 3.797 | 0.042 | 6-31+g(d) |
| | -0.729 | -6.798 | -3.763 | 6.069 | 1.167 | 6.798 | 0.729 | 6.069 | 3.807 | 0.044 | 6-311+g(d) |
| | -0.729 | -6.798 | -3.763 | 6.069 | 1.167 | 6.798 | 0.729 | 6.069 | 3.807 | 0.044 | 6-311++g(d) |
| $O^{-}$ | 2.895 | -9.672 | -3.388 | 12.567 | 0.457 | 9.672 | -2.895 | 12.567 | 3.722 | 0.334 | 6-31+g(d) |
| | 2.874 | -9.672 | -3.399 | 12.546 | 0.461 | 9.672 | -2.874 | 12.546 | 3.728 | 0.329 | 6-311+g(d) |
| | 2.874 | -9.672 | -3.399 | 12.546 | 0.461 | 9.672 | -2.874 | 12.546 | 3.728 | 0.329 | 6-311++g(d) |
| $F^{-}$ | 1.278 | -14.266 | -6.484 | 15.543 | 1.357 | 14.266 | -1.278 | 15.544 | 6.547 | 0.052 | 6-31+g(d) |
| | 1.245 | -14.409 | -6.582 | 15.654 | 1.357 | 14.409 | -1.245 | 15.654 | 6.631 | 0.049 | 6-311+g(d) |
| | 1.245 | -14.409 | -6.582 | 15.654 | 1.357 | 14.409 | -1.245 | 15.654 | 6.631 | 0.049 | 6-311++g(d) |
| $Ne^{-}$ | -7.852 | -16.322 | -12.087 | 8.470 | 8.624 | 16.322 | 7.852 | 8.470 | 15.727 | 3.639 | 6-31+g(d) |
| | -7.553 | -15.900 | -11.726 | 8.348 | 8.236 | 15.900 | 7.553 | 8.347 | 15.143 | 3.416 | 6-311+g(d) |
| | -7.553 | -15.900 | -11.726 | 8.348 | 8.236 | 15.900 | 7.553 | 8.347 | 15.143 | 3.416 | 6-311++g(d) |

(e)

| Ions | I (eV) | E (eV) | χ (eV) | η (eV) | ω (eV) | $\mu^+$ (eV) | $\mu^-$ (eV) | η (eV) | $\omega^+$ (eV) | $\omega^-$ (eV) | Basis sets |
|---|---|---|---|---|---|---|---|---|---|---|---|
| $Li^{2-}$ | -2.130 | -4.641 | -3.385 | 2.511 | 2.282 | 4.641 | 2.130 | 2.511 | 4.288 | 0.903 | 6-31+g(d) |
| | -2.383 | -4.168 | -3.276 | 1.785 | 3.005 | 4.168 | 2.383 | 1.785 | 4.866 | 1.590 | 6-311+g(d) |
| | -2.383 | -4.168 | -3.276 | 1.785 | 3.005 | 4.168 | 2.383 | 1.785 | 4.866 | 1.590 | 6-311++g(d) |
| $Be^{2-}$ | -4.541 | -6.435 | -5.488 | 1.895 | 7.949 | 6.435 | 4.541 | 1.894 | 10.930 | 5.442 | 6-31+g(d) |
| | -4.412 | -6.130 | -5.271 | 1.718 | 8.085 | 6.130 | 4.412 | 1.718 | 10.935 | 5.665 | 6-311+g(d) |
| | -4.412 | -6.130 | -5.271 | 1.718 | 8.085 | 6.130 | 4.412 | 1.718 | 10.935 | 5.665 | 6-311++g(d) |
| $B^{2-}$ | -4.911 | -9.334 | -7.122 | 4.423 | 5.735 | 9.334 | 4.911 | 4.423 | 9.849 | 2.727 | 6-31+g(d) |
| | -4.902 | -45.491 | -25.196 | 40.589 | 7.821 | 45.491 | 4.902 | 40.589 | 25.492 | 0.296 | 6-311+g(d) |
| | -4.902 | -45.491 | -25.196 | 40.589 | 7.821 | 45.491 | 4.902 | 40.589 | 25.492 | 0.296 | 6-311++g(d) |
| $C^{2-}$ | -7.334 | -9.431 | -8.382 | 2.097 | 16.755 | 9.431 | 7.334 | 2.097 | 21.208 | 12.826 | 6-31+g(d) |
| | -7.410 | -10.577 | -8.994 | 3.168 | 12.767 | 10.577 | 7.410 | 3.167 | 17.660 | 8.666 | 6-311+g(d) |
| | -7.410 | -10.577 | -8.994 | 3.168 | 12.767 | 10.577 | 7.410 | 3.167 | 17.660 | 8.666 | 6-311++g(d) |
| $N^{2-}$ | -6.794 | -14.316 | -10.555 | 7.522 | 7.405 | 14.316 | 6.794 | 7.522 | 13.623 | 3.068 | 6-31+g(d) |
| | -6.798 | -14.253 | -10.626 | 7.456 | 7.429 | 14.253 | 6.798 | 7.455 | 13.625 | 3.099 | 6-311+g(d) |
| | -6.798 | -14.253 | -10.626 | 7.456 | 7.429 | 14.253 | 6.798 | 7.455 | 16.625 | 3.099 | 6-311++g(d) |
| $O^{2-}$ | -9.672 | -16.889 | -13.280 | 7.217. | 12.218 | 16.889 | 9.672 | 7.217 | 19.760 | 6.480 | 6-31+g(d) |
| | -9.672 | -18.608 | -14.140 | 8.936 | 11.187 | 18.608 | 9.672 | 8.936 | 19.374 | 5.234 | 6-311+g(d) |
| | -9.672 | -18.608 | -14.140 | 8.936 | 11.187 | 18.608 | 9.672 | 8.936 | 19.374 | 5.234 | 6-311++g(d) |
| $F^{2-}$ | -14.266 | -21.508 | -17.887 | 7.242 | 22.089 | 21.508 | 14.266 | 7.242 | 31.937 | 14.051 | 6-31+g(d) |
| | -14.409 | -21.751 | -18.084 | 7.351 | 22.245 | 21.751 | 14.409 | 7.352 | 32.206 | 14.121 | 6-311+g(d) |
| | -14.409 | -21.751 | -18.084 | 7.351 | 44.489 | 21.751 | 14.409 | 7.352 | 32.206 | 14.121 | 6-311++g(d) |
| $Ne^{2-}$ | -16.322 | -23.684 | -20.003 | 7.362 | 27.176 | 23.684 | 16.322 | 7.362 | 38.097 | 18.095 | 6-31+g(d) |
| | -15.900 | -23.084 | -19.492 | 7.184 | 26.445 | 23.084 | 15.900 | 7.184 | 37.089 | 17.597 | 6-311+g(d) |
| | -15.900 | -23.084 | -19.492 | 7.184 | 26.445 | 23.084 | 15.900 | 7.184 | 37.089 | 17.597 | 6-311++g(d) |

**Table 4**: Ionization potential(I) , Electron affinity(E), Electronegativity($\chi$), Chemical hardness($\eta$), Electrophilicity($\omega$) and the values of ($\mu^+$, $\mu^-$ ,$\omega^+$ ,$\omega^-$ ) for Electroaccepting and Electrodonating processes of atoms and ions by using MP2 method

(a)

| **Atoms** | **I (eV)** | **E (eV)** | **$\chi$ (eV)** | **$\eta$ (eV)** | **$\omega$ (eV)** | **$\mu^+$ (eV)** | **$\mu^-$ (eV)** | **$\eta$ (eV)** | **$\omega^+$ (eV)** | **$\omega^-$ (eV)** | **Basis sets** |
|---|---|---|---|---|---|---|---|---|---|---|---|
| Li | 5.334 | -0.156 | 2.589 | 5.450 | 0.610 | 0.156 | -5.334 | 5.450 | 0.002 | 2.591 | 6-31+g(d) |
| | 5.338 | -0.124 | 2.607 | 5.462 | 0.622 | 0.124 | -5.338 | 5.462 | 0.001 | 2.609 | 6-311+g(d) |
| | 5.338 | -0.124 | 2.607 | 5.462 | 0.622 | 0.124 | -5.338 | 5.462 | 0.001 | 0.206 | 6-311++g(d) |
| Be | 8.006 | -0.708 | 3.649 | 8.713 | 0.764 | 0.708 | -8.006 | 8.714 | 0.029 | 3.678 | 6-31+g(d) |
| | 8.047 | -0.703 | 3.672 | 8.751 | 0.771 | 0.703 | -8.047 | 8.750 | 0.028 | 3.700 | 6-311+g(d) |
| | 8.047 | -0.703 | 3.672 | 8.750 | 0.771 | 0.703 | -8.047 | 8.750 | 0.028 | 3.700 | 6-311++g(d) |
| B | 7.947 | -1.542 | 3.203 | 9.490 | 0.541 | 1.542 | -7.947 | 9.489 | 0.125 | 3.328 | 6-31+g(d) |
| | 8.022 | -1.574 | 3.224 | 9.596 | 0.542 | 1.574 | -8.022 | 9.596 | 0.129 | 3.353 | 6-311+g(d) |
| | 8.022 | -1.574 | 3.224 | 9.596 | 0.542 | 1.574 | -8.022 | 9.596 | 0.129 | 3.353 | 6-311++g(d) |
| C | 8.357 | 0.700 | 4.528 | 7.657 | 1.339 | -0.700 | -8.357 | 7.657 | 0.032 | 4.561 | 6-31+g(d) |
| | 8.378 | 0.700 | 4.536 | 7.684 | 1.339 | -0.700 | -8.378 | 7.678 | 0.031 | 4.568 | 6-311+g(d) |
| | 8.378 | 0.700 | 4.536 | 7.684 | 1.339 | -0.700 | -8.378 | 7.678 | 0.031 | 4.568 | 6-311++g(d) |
| N | 13.621 | -0.716 | 6.452 | 14.337 | 1.452 | 0.716 | -13.621 | 14.337 | 0.018 | 6.471 | 6-31+g(d) |
| | 13.586 | -0.729 | 6.428 | 14.315 | 1.443 | 0.729 | -13.586 | 14.315 | 0.018 | 6.447 | 6-311+g(d) |
| | 13.585 | -0.729 | 6.428 | 14.315 | 1.443 | 0.729 | -13.585 | 14.315 | 0.018 | 6.447 | 6-311++g(d) |
| O | 13.617 | 2.895 | 8.256 | 10.722 | 3.179 | -2.895 | -13.617 | 10.722 | 0.391 | 8.647 | 6-31+g(d) |
| | 13.576 | 2.874 | 8.225 | 10.702 | 3.161 | -2.874 | -13.576 | 10.702 | 0.386 | 8.611 | 6-311+g(d) |
| | 13.576 | 2.874 | 8.225 | 10.702 | 3.161 | -2.874 | -13.576 | 10.702 | 0.386 | 8.611 | 6-311++g(d) |
| F | 20.106 | 1.277 | 10.691 | 18.828 | 3.035 | -1.277 | -20.106 | 18.827 | 0.043 | 10.735 | 6-31+g(d) |
| | 20.092 | 1.245 | 10.668 | 18.846 | 3.019 | -1.245 | -20.092 | 18.847 | 0.041 | 10.710 | 6-311+g(d) |
| | 20.092 | 1.245 | 10.668 | 18.846 | 3.019 | -1.245 | -20.092 | 18.847 | 0.041 | 10.710 | 6-311++g(d) |
| Ne | 19.891 | -7.852 | 6.019 | 27.743 | 0.653 | 7.852 | -19.891 | 27.743 | 1.111 | 7.131 | 6-31+g(d) |
| | 19.794 | -7.553 | 6.121 | 27.346 | 0.685 | 7.553 | -19.794 | 27.347 | 1.043 | 7.164 | 6-311+g(d) |
| | 19.794 | -7.553 | 6.121 | 27.346 | 0.685 | 7.553 | -19.794 | 27.347 | 1.043 | 7.164 | 6-311++g(d) |

(b)

| **Ions** | **I (eV)** | **E (eV)** | **$\chi$ (eV)** | **$\eta$ (eV)** | **$\omega$ (eV)** | **$\mu^+$ (eV)** | **$\mu^-$ (eV)** | **$\eta$ (eV)** | **$\omega^+$ (eV)** | **$\omega^-$ (eV)** | **Basis sets** |
|---|---|---|---|---|---|---|---|---|---|---|---|
| $Li^+$ | 75.644 | 5.334 | 40.489 | 70.310 | 11.658 | -5.334 | -75.644 | 70.310 | 0.202 | 40.692 | 6-31+g(d) |
| | 74.612 | 5.338 | 39.975 | 69.273 | 11.534 | -5.338 | -79.951 | 69.273 | 0.206 | 40.181 | 6-311+g(d) |
| | 74.612 | 5.338 | 39.975 | 69.273 | 11.534 | -5.338 | -79.951 | 69.273 | 0.206 | 40.181 | 6-311++g(d) |
| $Be^+$ | 18.115 | 8.006 | 13.060 | 10.110 | 8.436 | -8.006 | -18.115 | 10.109 | 3.170 | 16.230 | 6-31+g(d) |
| | 18.118 | 8.047 | 13.083 | 10.071 | 8.498 | -8.047 | -18.118 | 10.979 | 3.215 | 16.298 | 6-311+g(d) |
| | 18.118 | 8.047 | 13.083 | 10.071 | 8.498 | -8.047 | -18.118 | 10.071 | 3.215 | 16.298 | 6-311++g(d) |
| $B^+$ | 23.451 | 7.948 | 15.699 | 15.503 | 7.949 | -7.948 | -23.451 | 15.503 | 2.037 | 17.736 | 6-31+g(d) |
| | 23.442 | 8.022 | 15.732 | 15.420 | 8.025 | -8.022 | -23.442 | 15.422 | 2.086 | 17.818 | 6-311+g(d) |
| | 23.442 | 8.022 | 15.732 | 15.420 | 8.025 | -8.022 | -23.442 | 15.422 | 2.086 | 17.818 | 6-311++g(d) |
| $C^+$ | 24.137 | 8.357 | 16.247 | 15.780 | 8.364 | -8.357 | -24.137 | 15.780 | 2.213 | 18.460 | 6-31+g(d) |
| | 24.147 | 8.378 | 16.263 | 15.769 | 8.386 | -8.378 | -24.147 | 15.768 | 2.226 | 18.488 | 6-311+g(d) |
| | 24.147 | 8.378 | 16.263 | 15.769 | 8.386 | -8.378 | -24.147 | 15.768 | 2.226 | 18.488 | 6-311++g(d) |
| $N^+$ | 25.842 | 13.621 | 19.732 | 12.221 | 15.929 | -13.621 | -25.842 | 12.221 | 7.591 | 27.323 | 6-31+g(d) |
| | 25.783 | 13.586 | 19.684 | 12.197 | 15.884 | -13.586 | -25.783 | 12.783 | 7.566 | 27.251 | 6-311+g(d) |
| | 25.783 | 13.586 | 19.685 | 12.197 | 15.884 | -13.586 | -25.783 | 12.783 | 7.566 | 27.251 | 6-311++g(d) |
| $O^+$ | 33.926 | 13.617 | 23.771 | 20.308 | 13.912 | -13.617 | -33.926 | 20.309 | 4.565 | 28.336 | 6-31+g(d) |
| | 33.878 | 13.576 | 23.727 | 20.302 | 13.865 | -13.576 | -33.878 | 20.302 | 4.539 | 28.266 | 6-311+g(d) |
| | 33.878 | 13.576 | 23.727 | 20.302 | 13.865 | -13.576 | -33.878 | 20.302 | 4.539 | 28.266 | 6-311++g(d) |
| $F^+$ | 35.262 | 20.106 | 27.684 | 15.157 | 25.293 | -20.106 | -35.262 | 15.156 | 13.335 | 41.019 | 6-31+g(d) |
| | 35.185 | 20.092 | 27.638 | 15.094 | 25.305 | -20.092 | -35.185 | 15.094 | 13.372 | 41.011 | 6-311+g(d) |
| | 35.185 | 20.092 | 27.638 | 15.094 | 25.305 | -20.092 | -35.185 | 15.094 | 13.372 | 41.011 | 6-311++g(d) |
| $Ne^+$ | 44.591 | 19.891 | 32.241 | 24.700 | 21.042 | -19.891 | -44.591 | 24.700 | 8.009 | 40.249 | 6-31+g(d) |
| | 44.577 | 19.794 | 32.186 | 24.784 | 20.899 | -19.794 | -44.577 | 24.777 | 7.904 | 40.090 | 6-311+g(d) |
| | 44.577 | 19.794 | 32.186 | 24.784 | 20.899 | -19.794 | -44.577 | 24.777 | 7.904 | 40.090 | 6-311++g(d) |

(c)

| Ions | I (eV) | E (eV) | χ (eV) | η (eV) | ω (eV) | $\mu^+$ (eV) | $\mu^-$ (eV) | η (eV) | $\omega^+$ (eV) | $\omega^-$ (eV) | Basis sets |
|---|---|---|---|---|---|---|---|---|---|---|---|
| $Li^{2+}$ | 121.245 | 75.644 | 98.445 | 45.601 | 106.263 | -75.644 | -121.245 | 45.601 | 62.741 | 161.186 | 6-31+g(d) |
| | 122.286 | 74.612 | 98.449 | 47.674 | 101.650 | -74.612 | -122.286 | 47.674 | 58.385 | 156.833 | 6-311+g(d) |
| | 122.286 | 74.116 | 98.449 | 47.674 | 101.650 | -74.116 | -122.286 | 47.674 | 58.385 | 156.833 | 6-311++g(d) |
| $Be^{2+}$ | 153.250 | 18.115 | 85.682 | 135.134 | 27.164 | -18.115 | -153.250 | 135.135 | 1.214 | 86.897 | 6-31+g(d) |
| | 152.879 | 18.118 | 85.498 | 134.761 | 27.164 | -18.118 | -152.879 | 134.761 | 1.218 | 86.716 | 6-311+g(d) |
| | 152.879 | 18.118 | 85.498 | 134.761 | 27.164 | -18.118 | -152.879 | 134.761 | 1.218 | 86.716 | 6-311++g(d) |
| $B^{2+}$ | 37.781 | 23.451 | 30.616 | 14.329 | 32.705 | -23.451 | -37.781 | 14.330 | 19.188 | 49.804 | 6-31+g(d) |
| | 37.801 | 23.442 | 30.622 | 14.359 | 32.651 | -23.442 | -37.801 | 14.359 | 19.136 | 49.757 | 6-311+g(d) |
| | 37.801 | 23.442 | 30.622 | 14.359 | 32.651 | -23.442 | -37.801 | 14.359 | 19.136 | 49.757 | 6-311++g(d) |
| $C^{2+}$ | 45.762 | 24.137 | 34.950 | 21.624 | 28.243 | -24.137 | -45.762 | 21.625 | 13.472 | 48.421 | 6-31+g(d) |
| | 45.802 | 24.147 | 34.975 | 21.655 | 28.243 | -24.147 | -45.802 | 21.655 | 13.463 | 48.738 | 6-311+g(d) |
| | 45.802 | 24.147 | 34.975 | 21.655 | 28.243 | -24.147 | -45.802 | 21.655 | 13.463 | 48.738 | 6-311++g(d) |
| $N^{2+}$ | 47.271 | 25.842 | 36.557 | 21.429 | 31.182 | -25.842 | -47.271 | 21.429 | 15.592 | 52.139 | 6-31+g(d) |
| | 47.310 | 25.783 | 36.547 | 21.526 | 31.023 | -25.783 | -47.310 | 21.527 | 15.440 | 51.987 | 6-311+g(d) |
| | 47.310 | 25.783 | 36.546 | 21.526 | 31.023 | -25.783 | -47.310 | 21.527 | 15.440 | 51.987 | 6-311++g(d) |
| $O^{2+}$ | 50.378 | 33.925 | 42.152 | 16.453 | 53.997 | -33.925 | -50.378 | 16.453 | 34.978 | 77.129 | 6-31+g(d) |
| | 50.277 | 33.878 | 42.078 | 16.399 | 53.984 | -33.878 | -50.277 | 16.399 | 34.995 | 77.073 | 6-311+g(d) |
| | 50.277 | 33.878 | 42.078 | 16.399 | 53.984 | -33.878 | -50.277 | 16.399 | 34.995 | 77.073 | 6-311++g(d) |
| $F^{2+}$ | 61.241 | 35.262 | 48.252 | 25.978 | 44.811 | -35.262 | -61.241 | 25.979 | 23.932 | 72.184 | 6-31+g(d) |
| | 61.219 | 35.185 | 48.202 | 26.033 | 44.624 | -35.185 | -61.219 | 26.034 | 23.777 | 71.980 | 6-311+g(d) |
| | 61.219 | 35.185 | 48.202 | 26.033 | 44.624 | -35.185 | -61.219 | 26.034 | 23.777 | 71.980 | 6-311++g(d) |
| $Ne^{2+}$ | 63.935 | 44.591 | 54.263 | 19.343 | 76.109 | -44.591 | -63.935 | 19.344 | 51.396 | 105.659 | 6-31+g(d) |
| | 63.864 | 44.577 | 54.221 | 19.186 | 76.216 | -44.577 | -63.864 | 19.187 | 51.516 | 105.737 | 6-311+g(d) |
| | 63.864 | 44.577 | 54.221 | 19.186 | 76.216 | -44.577 | -63.864 | 19.187 | 51.516 | 105.737 | 6-311++g(d) |

(d)

| Ions | I (eV) | E (eV) | χ (eV) | η (eV) | ω (eV) | $\mu^+$ (eV) | $\mu^-$ (eV) | η (eV) | $\omega^+$ (eV) | $\omega^-$ (eV) | Basis sets |
|---|---|---|---|---|---|---|---|---|---|---|---|
| $Li^-$ | -0.156 | -2.128 | -1.142 | 1.972 | 0.330 | 2.128 | 0.156 | 1.972 | 1.148 | 0.006 | 6-31+g(d) |
| | -0.124 | -2.383 | -1.253 | 2.259 | 0.347 | 2.383 | 0.124 | 2.259 | 1.257 | 0.003 | 6-311+g(d) |
| | -0.124 | -2.383 | -1.253 | 2.259 | 0.347 | 2.383 | 0.124 | 2.259 | 1.257 | 0.003 | 6-311++g(d) |
| $Be^-$ | -0.708 | -4.541 | -2.624 | 3.833 | 0.898 | 4.541 | 0.708 | 3.833 | 2.690 | 0.065 | 6-31+g(d) |
| | -0.703 | -4.412 | -2.557 | 3.708 | 0.882 | 4.412 | 0.703 | 3.709 | 2.624 | 0.067 | 6-311+g(d) |
| | -0.703 | -2.383 | -2.557 | 3.708 | 0.882 | 2.383 | 0.703 | 3.709 | 2.624 | 0.067 | 6-311++g(d) |
| $B^-$ | -1.542 | -4.911 | -3.226 | 3.369 | 1.545 | 4.911 | 1.542 | 3.369 | 3.579 | 0.353 | 6-31+g(d) |
| | -1.574 | -4.902 | -3.238 | 3.328 | 1.575 | 4.902 | 1.574 | 3.328 | 3.610 | 0.372 | 6-311+g(d) |
| | -1.574 | -4.902 | -3.238 | 3.328 | 1.575 | 4.902 | 1.574 | 3.328 | 3.610 | 0.372 | 6-311++g(d) |
| $C^-$ | 0.700 | -7.334 | -3.317 | 8.034 | 0.685 | 7.334 | -0.700 | 8.034 | 3.348 | 0.030 | 6-31+g(d) |
| | 0.700 | -7.410 | -3.358 | 8.104 | 0.696 | 7.410 | -0.700 | 8.104 | 3.387 | 0.030 | 6-311+g(d) |
| | 0.700 | -7.410 | -3.358 | 8.104 | 0.696 | 7.410 | -0.700 | 8.104 | 3.387 | 0.030 | 6-311++g(d) |
| $N^-$ | -0.716 | -6.798 | -3.755 | 6.078 | 1.160 | 6.798 | 0.716 | 6.082 | 3.797 | 0.042 | 6-31+g(d) |
| | -0.729 | -6.798 | -3.764 | 6.068 | 1.167 | 6.798 | 0.729 | 6.069 | 3.807 | 0.044 | 6-311+g(d) |
| | -0.729 | -6.798 | -3.764 | 6.068 | 1.167 | 6.798 | 0.729 | 6.069 | 3.807 | 0.044 | 6-311++g(d) |
| $O^-$ | 2.895 | -9.672 | -3.388 | 12.567 | 0.457 | 9.672 | -2.895 | 12.567 | 3.722 | 0.334 | 6-31+g(d) |
| | 2.874 | -9.672 | -3.399 | 12.547 | 0.460 | 9.672 | -2.874 | 12.546 | 3.728 | 0.329 | 6-311+g(d) |
| | 2.874 | -9.672 | -3.399 | 12.547 | 0.460 | 9.672 | -2.874 | 12.546 | 3.728 | 0.329 | 6-311++g(d) |
| $F^-$ | 1.277 | -14.266 | -6.484 | 15.543 | 1.357 | 14.266 | -1.277 | 15.543 | 6.547 | 0.052 | 6-31+g(d) |
| | 1.245 | -14.409 | -6.582 | 15.654 | 1.384 | 14.409 | -1.245 | 15.654 | 6.631 | 0.049 | 6-311+g(d) |
| | 1.245 | -14.409 | -6.582 | 15.654 | 1.384 | 14.409 | -1.245 | 15.654 | 6.631 | 0.049 | 6-311++g(d) |
| $Ne^-$ | -7.852 | -16.322 | -12.087 | 8.470 | 8.624 | 16.322 | 7.852 | 8.470 | 15.726 | 3.639 | 6-31+g(d) |
| | -7.553 | -15.900 | -11.726 | 8.348 | 8.236 | 15.900 | 7.553 | 8.347 | 15.143 | 3.416 | 6-311+g(d) |
| | -7.553 | -15.901 | -11.726 | 8.348 | 8.236 | 15.901 | 7.553 | 8.348 | 15.143 | 3.416 | 6-311++g(d) |

(e)

| Ions | I (eV) | E (eV) | χ (eV) | η (eV) | ω (eV) | $\mu^+$ (eV) | $\mu^-$ (eV) | η (eV) | $\omega^+$ (eV) | $\omega^-$ (eV) | Basis sets |
|---|---|---|---|---|---|---|---|---|---|---|---|
| $Li^{2-}$ | -2.127 | -4.636 | -3.382 | 2.508 | 2.280 | 4.636 | 2.127 | 2.509 | 4.284 | 0.902 | 6-31+g(d) |
| | -2.383 | -4.168 | -3.276 | 1.785 | 3.005 | 4.168 | 2.383 | 1.785 | 4.866 | 1.590 | 6-311+g(d) |
| | -2.383 | -4.168 | -3.276 | 1.785 | 3.005 | 4.168 | 2.383 | 1.785 | 4.866 | 1.590 | 6-311++g(d) |
| $Be^{2-}$ | -4.541 | -6.435 | -5.488 | 1.895 | 7.949 | 6.435 | 4.541 | 1.894 | 10.930 | 5.442 | 6-31+g(d) |
| | -4.412 | -6.130 | -5.271 | 1.718 | 8.085 | 6.130 | 4.412 | 1.718 | 10.935 | 5.665 | 6-311+g(d) |
| | -4.412 | -6.130 | -5.271 | 1.718 | 8.085 | 6.130 | 4.412 | 1.718 | 10.935 | 5.665 | 6-311++g(d) |
| $B^{2-}$ | -4.911 | -9.334 | -7.122 | 4.422 | 5.735 | 9.334 | 4.911 | 4.423 | 9.849 | 2.727 | 6-31+g(d) |
| | -4.902 | -9.252 | -7.077 | 4.350 | 5.757 | 9.252 | 4.902 | 4.350 | 9.839 | 2.762 | 6-311+g(d) |
| | -4.902 | -9.252 | -7.077 | 4.350 | 5.757 | 9.252 | 4.902 | 4.350 | 9.839 | 2.762 | 6-311++g(d) |
| $C^{2-}$ | -7.334 | -9.431 | -8.382 | 2.097 | 16.755 | 9.431 | 7.334 | 2.097 | 21.209 | 12.826 | 6-31+g(d) |
| | -7.410 | -10.286 | -8.848 | 2.876 | 13.608 | 10.286 | 7.410 | 2.876 | 18.392 | 9.545 | 6-311+g(d) |
| | -7.410 | -10.286 | -8.848 | 2.876 | 13.608 | 10.286 | 7.410 | 2.876 | 18.392 | 9.545 | 6-311++g(d) |
| $N^{2-}$ | -6.794 | -14.316 | -10.555 | 7.522 | 7.405 | 14.316 | 6.794 | 7.522 | 13.623 | 3.068 | 6-31+g(d) |
| | -6.798 | -14.253 | -10.626 | 7.455 | 7.430 | 14.253 | 6.798 | 7.455 | 13.625 | 3.099 | 6-311+g(d) |
| | -6.798 | -14.253 | -10.626 | 7.455 | 7.430 | 14.253 | 6.798 | 7.455 | 13.625 | 3.099 | 6-311++g(d) |
| $O^{2-}$ | -9.671 | -16.888 | -13.280 | 7.216 | 12.218 | 16.888 | 9.671 | 7.217 | 19.760 | 6.481 | 6-31+g(d) |
| | -9.673 | -17.001 | -13.337 | 7.328 | 12.136 | 17.001 | 9.673 | 7.328 | 19.721 | 6.384 | 6-311+g(d) |
| | -9.673 | -17.001 | -13.337 | 7.328 | 12.136 | 17.001 | 9.673 | 7.328 | 19.721 | 6.384 | 6-311++g(d) |
| $F^{2-}$ | -14.266 | -21.508 | -17.887 | 7.242 | 22.089 | 21.508 | 14.266 | 7.242 | 31.938 | 14.051 | 6-31+g(d) |
| | -14.409 | -21.761 | -18.085 | 7.352 | 22.242 | 21.761 | 14.409 | 7.352 | 32.204 | 14.119 | 6-311+g(d) |
| | -14.409 | -21.761 | -18.085 | 7.352 | 22.242 | 21.761 | 14.409 | 7.352 | 32.204 | 14.119 | 6-311++g(d) |
| $Ne^{2-}$ | -16.322 | -23.684 | -20.003 | 7.362 | 27.176 | 23.684 | 16.322 | 7.362 | 38.098 | 18.095 | 6-31+g(d) |
| | -15.901 | -23.084 | -19.492 | 7.184 | 26.446 | 23.084 | 15.901 | 7.183 | 37.090 | 17.597 | 6-311+g(d) |
| | -15.901 | -23.084 | -19.492 | 7.184 | 26.446 | 23.084 | 15.901 | 7.183 | 37.090 | 17.597 | 6-311++g(d) |

**Table 5**: Ionization potential(I) , Electron affinity(E), Electronegativity($\chi$), Chemical hardness($\eta$) , Electrophilicity($\omega$) and the values of ($\mu^+$, $\mu^-$ ,$\omega^+$ ,$\omega^-$ ) for Electroaccepting and Electrodonating processes of atoms and ions by using B3LYP method

(a)

| Atoms | I (eV) | E (eV) | $\chi$ (eV) | $\eta$ (eV) | $\omega$ (eV) | $\mu^+$ (eV) | $\mu^-$ (eV) | $\eta$ (eV) | $\omega^+$ (eV) | $\omega^-$ (eV) | Basis sets |
|---|---|---|---|---|---|---|---|---|---|---|---|
| Li | 5.622 | 0.548 | 3.085 | 5.074 | 0.938 | -0.548 | -5.622 | 5.074 | 0.029 | 3.114 | 6-31+g(d) |
| | 5.617 | 0.558 | 3.087 | 5.059 | 0.942 | -0.558 | -5.617 | 5.059 | 0.031 | 3.118 | 6-311+g(d) |
| | 5.617 | 0.558 | 3.087 | 5.059 | 0.942 | -0.558 | -5.617 | 5.059 | 0.031 | 3.118 | 6-311++g(d) |
| Be | 9.110 | -0.232 | 4.439 | 9.342 | 1.055 | 0.232 | -9.110 | 9.342 | 0.003 | 4.442 | 6-31+g(d) |
| | 9.117 | -0.227 | 4.445 | 9.344 | 1.057 | 0.227 | -9.117 | 9.344 | 0.003 | 4.448 | 6-311+g(d) |
| | 9.117 | -0.227 | 4.445 | 9.344 | 1.057 | 0.227 | -9.117 | 9.344 | 0.003 | 4.448 | 6-311++g(d) |
| B | 8.679 | -0.326 | 4.177 | 9.005 | 0.969 | 0.326 | -8.679 | 9.001 | 0.006 | 4.183 | 6-31+g(d) |
| | 8.728 | -0.350 | 4.189 | 9.079 | 0.966 | 0.350 | -8.728 | 9.078 | 0.007 | 4.196 | 6-311+g(d) |
| | 8.728 | -0.350 | 4.189 | 9.078 | 0.966 | 0.350 | -8.728 | 9.078 | 0.007 | 4.196 | 6-311++g(d) |
| C | 9.728 | 1.672 | 5.700 | 8.055 | 2.016 | -1.672 | -9.728 | 8.056 | 0.174 | 5.873 | 6-31+g(d) |
| | 9.770 | 1.642 | 5.706 | 8.128 | 2.003 | -1.642 | -9.771 | 8.128 | 0.166 | 5.872 | 6-311+g(d) |
| | 9.771 | 1.642 | 5.706 | 8.128 | 2.003 | -1.642 | -9.771 | 8.128 | 0.166 | 5.872 | 6-311++g(d) |
| N | 14.635 | 1.072 | 7.853 | 13.562 | 2.274 | -1.0721 | -14.635 | 13.562 | 0.042 | 7.895 | 6-31+g(d) |
| | 14.603 | 1.058 | 7.831 | 13.544 | 2.264 | -1.058 | -14.603 | 13.544 | 0.041 | 7.872 | 6-311+g(d) |
| | 14.603 | 1.058 | 7.831 | 13.544 | 2.264 | -1.058 | -14.603 | 13.544 | 0.041 | 7.872 | 6-311++g(d) |
| O | 15.317 | 4.377 | 9.847 | 10.940 | 4.432 | -4.377 | -15.317 | 10.940 | 0.876 | 10.723 | 6-31+g(d) |
| | 15.293 | 4.357 | 9.824 | 10.936 | 4.413 | -4.357 | -15.293 | 10.936 | 0.868 | 10.692 | 6-311+g(d) |
| | 15.293 | 4.357 | 9.825 | 10.936 | 4.413 | -4.357 | -15.293 | 10.936 | 0.868 | 10.692 | 6-311++g(d) |
| F | 21.406 | 3.513 | 12.460 | 17.893 | 4.338 | -3.513 | -21.406 | 17.893 | 0.345 | 12.804 | 6-31+g(d) |
| | 21.397 | 3.486 | 12.442 | 17.911 | 4.321 | -3.486 | -21.397 | 17.911 | 0.339 | 12.781 | 6-311+g(d) |
| | 21.397 | 3.486 | 12.442 | 17.911 | 4.321 | -3.486 | -21.397 | 17.911 | 0.339 | 12.781 | 6-311++g(d) |
| Ne | 21.844 | -6.833 | 7.505 | 28.677 | 0.982 | 6.833 | -21.844 | 28.677 | 0.814 | 8.319 | 6-31+g(d) |
| | 21.802 | -6.584 | 7.609 | 28.385 | 1.020 | 6.583 | -21.802 | 28.385 | 0.764 | 8.372 | 6-311+g(d) |
| | 21.802 | -6.583 | 7.609 | 28.385 | 1.019 | 6.583 | -21.802 | 28.385 | 0.764 | 8.372 | 6-311++g(d) |

(b)

| Ions | I (eV) | E (eV) | $\chi$ (eV) | $\eta$ (eV) | $\omega$ (eV) | $\mu^+$ (eV) | $\mu^-$ (eV) | $\eta$ (eV) | $\omega^+$ (eV) | $\omega^-$ (eV) | Basis sets |
|---|---|---|---|---|---|---|---|---|---|---|---|
| $Li^+$ | 77.073 | 5.622 | 41.347 | 71.451 | 11.964 | -5.622 | -77.073 | 71.451 | 0.221 | 41.569 | 6-31+g(d) |
| | 76.047 | 5.617 | 40.832 | 70.430 | 11.836 | -5.617 | -76.047 | 70.430 | 0.224 | 41.056 | 6-311+g(d) |
| | 76.047 | 5.617 | 40.832 | 70.430 | 11.836 | -5.617 | -76.047 | 70.430 | 0.224 | 41.056 | 6-311++g(d) |
| $Be^2$ | 18.604 | 9.110 | 13.857 | 9.494 | 10.112 | -9.110 | -18.604 | 9.494 | 4.370 | 18.227 | 6-31+g(d) |
| | 18.595 | 9.117 | 13.856 | 9.478 | 10.129 | -9.117 | -18.595 | 9.478 | 4.385 | 18.242 | 6-311+g(d) |
| | 18.595 | 9.117 | 13.856 | 9.478 | 10.129 | -9.117 | -18.595 | 9.478 | 4.385 | 18.242 | 6-311++g(d) |
| $B^+$ | 24.767 | 8.679 | 16.723 | 16.087 | 8.692 | -8.679 | -24.767 | 16.087 | 2.341 | 19.065 | 6-31+g(d) |
| | 24.739 | 8.728 | 16.734 | 16.011 | 8.744 | -8.728 | -24.739 | 16.011 | 2.379 | 19.113 | 6-311+g(d) |
| | 24.739 | 8.728 | 16.734 | 16.011 | 8.744 | -8.728 | -24.739 | 16.011 | 2.379 | 19.113 | 6-311++g(d) |
| $C^+$ | 25.023 | 9.728 | 17.375 | 15.296 | 9.869 | -9.728 | -25.023 | 15.296 | 3.093 | 20.468 | 6-31+g(d) |
| | 25.032 | 9.770 | 17.401 | 15.261 | 9.921 | -9.771 | -25.032 | 15.261 | 3.128 | 20.529 | 6-311+g(d) |
| | 25.032 | 9.771 | 17.401 | 15.261 | 9.921 | -9.771 | -25.032 | 15.261 | 3.128 | 20.529 | 6-311++g(d) |
| $N^+$ | 27.379 | 14.635 | 21.007 | 12.745 | 17.313 | -14.635 | -27.379 | 12.745 | 8.402 | 29.409 | 6-31+g(d) |
| | 27.315 | 14.603 | 20.959 | 12.713 | 17.278 | -14.603 | -27.315 | 12.712 | 8.387 | 29.346 | 6-311+g(d) |
| | 27.315 | 14.603 | 20.959 | 12.713 | 17.278 | -14.603 | -27.315 | 12.712 | 8.387 | 29.346 | 6-311++g(d) |
| $O^+$ | 35.014 | 15.317 | 25.165 | 19.697 | 16.076 | -15.317 | -35.014 | 19.697 | 5.955 | 31.121 | 6-31+g(d) |
| | 34.941 | 15.293 | 25.117 | 19.648 | 16.054 | -15.293 | -34.941 | 19.648 | 5.951 | 31.068 | 6-311+g(d) |
| | 34.941 | 15.293 | 25.117 | 19.648 | 16.054 | -15.293 | -34.941 | 19.648 | 5.951 | 31.068 | 6-311++g(d) |
| $F^+$ | 36.985 | 21.406 | 29.196 | 15.578 | 27.357 | -21.406 | -36.985 | 15.578 | 14.707 | 43.903 | 6-31+g(d) |
| | 36.913 | 21.397 | 29.155 | 15.516 | 27.392 | -21.397 | -36.913 | 15.516 | 14.754 | 43.910 | 6-311+g(d) |
| | 36.913 | 21.397 | 29.155 | 15.516 | 27.392 | -21.397 | -36.913 | 15.516 | 14.754 | 43.910 | 6-311++g(d) |
| $Ne^+$ | 45.858 | 21.844 | 33.851 | 24.015 | 23.858 | -21.844 | -45.858 | 24.015 | 9.934 | 43.785 | 6-31+g(d) |
| | 45.806 | 21.802 | 33.804 | 24.004 | 23.802 | -21.802 | -45.806 | 24.004 | 9.901 | 43.704 | 6-311+g(d) |
| | 45.806 | 21.802 | 33.804 | 24.004 | 23.802 | -21.802 | -45.806 | 24.004 | 9.901 | 43.704 | 6-311++g(d) |

(c)

| Ions | I (eV) | E (eV) | χ (eV) | η (eV) | ω (eV) | $\mu^+$ (eV) | $\mu^-$ (eV) | η (eV) | $\omega^+$ (eV) | $\omega^-$ (eV) | Basis sets |
|---|---|---|---|---|---|---|---|---|---|---|---|
| $Li^{2+}$ | 121.151 | 77.073 | 99.112 | 44.078 | 111.428 | -77.073 | -121.151 | 44.078 | 67.382 | 166.494 | 6-31+g(d) |
| | 122.186 | 76.047 | 99.116 | 46.139 | 106.461 | -76.047 | -122.186 | 46.139 | 62.670 | 161.786 | 6-311+g(d) |
| | 122.186 | 76.047 | 99.116 | 46.139 | 106.461 | -76.047 | -122.186 | 46.139 | 62.670 | 161.786 | 6-311++g(d) |
| $Be^{2+}$ | 154.717 | 18.604 | 86.660 | 136.113 | 27.587 | -18.604 | -154.717 | 136.113 | 1.271 | 87.932 | 6-31+g(d) |
| | 154.254 | 18.596 | 86.425 | 135.659 | 27.529 | -18.595 | -154.254 | 135.659 | 1.274 | 87.699 | 6-311+g(d) |
| | 154.254 | 18.595 | 86.425 | 135.659 | 27.529 | -18.595 | -154.254 | 135.659 | 1.274 | 87.699 | 6-311++g(d) |
| $B^{2+}$ | 38.425 | 24.767 | 31.5963 | 13.658 | 36.546 | -24.767 | -38.425 | 12.658 | 22.456 | 54.052 | 6-31+g(d) |
| | 38.447 | 24.739 | 31.593 | 13.708 | 36.406 | -24.739 | -38.447 | 13.708 | 22.323 | 53.916 | 6-311+g(d) |
| | 38.447 | 24.739 | 31.593 | 13.708 | 36.406 | -24.739 | -38.447 | 13.708 | 22.323 | 53.916 | 6-311++g(d) |
| $C^{2+}$ | 47.220 | 25.023 | 36.122 | 22.197 | 29.391 | -25.023 | -47.220 | 22.197 | 14.105 | 50.226 | 6-31+g(d) |
| | 47.254 | 25.032 | 36.143 | 22.223 | 29.392 | -25.032 | -47.254 | 22.223 | 14.098 | 50.241 | 6-311+g(d) |
| | 47.254 | 25.032 | 36.143 | 22.223 | 29.392 | -25.032 | -47.254 | 22.223 | 14.098 | 50.241 | 6-311++g(d) |
| $N^{2+}$ | 48.267 | 27.379 | 37.823 | 20.888 | 34.245 | -27.379 | -48.267 | 20.888 | 17.944 | 55.768 | 6-31+g(d) |
| | 48.333 | 27.315 | 37.824 | 21.018 | 34.034 | -27.315 | -48.333 | 21.018 | 17.749 | 55.574 | 6-311+g(d) |
| | 48.333 | 27.315 | 37.824 | 21.018 | 34.034 | -27.315 | -48.333 | 21.018 | 17.749 | 55.574 | 6-311++g(d) |
| $O^{2+}$ | 52.017 | 35.014 | 43.515 | 17.003 | 55.685 | -35.014 | -52.017 | 17.003 | 36.043 | 79.568 | 6-31+g(d) |
| | 51.917 | 34.941 | 43.429 | 16.976 | 55.550 | -34.941 | -51.917 | 16.976 | 35.958 | 79.387 | 6-311+g(d) |
| | 51.917 | 34.941 | 43.429 | 16.976 | 55.550 | -34.941 | -51.917 | 16.976 | 35.958 | 79.387 | 6-311++g(d) |
| $F^{2+}$ | 62.398 | 36.985 | 49.692 | 25.414 | 48.582 | -36.985 | -62.398 | 25.414 | 26.912 | 76.604 | 6-31+g(d) |
| | 62.318 | 36.913 | 49.616 | 25.405 | 48.450 | -36.913 | -62.318 | 25.405 | 26.818 | 76.434 | 6-311+g(d) |
| | 62.318 | 36.913 | 49.616 | 25.405 | 48.450 | -36.913 | -62.318 | 25.405 | 26.818 | 76.434 | 6-311++g(d) |
| $Ne^{2+}$ | 65.695 | 45.858 | 55.777 | 19.837 | 78.415 | -45.858 | -65.695 | 19.837 | 53.006 | 108.783 | 6-31+g(d) |
| | 65.605 | 45.806 | 55.705 | 19.779 | 78.364 | -45.806 | -65.605 | 19.799 | 52.986 | 108.691 | 6-311+g(d) |
| | 65.605 | 45.806 | 55.705 | 16.799 | 78.364 | -45.806 | -65.605 | 19.799 | 52.986 | 108.691 | 6-311++g(d) |

(d)

| Ions | I (eV) | E (eV) | χ (eV) | η (eV) | ω (eV) | $\mu^+$ (eV) | $\mu^-$ (eV) | η (eV) | $\omega^+$ (eV) | $\omega^-$ (eV) | Basis sets |
|---|---|---|---|---|---|---|---|---|---|---|---|
| $Li^-$ | 0.548 | -2.190 | -0.821 | 2.738 | 0.123 | 2.190 | -0.548 | 2.738 | 0.876 | 0.055 | 6-31+g(d) |
| | 0.558 | -2.102 | -0.772 | 2.660 | 0.112 | 2.102 | -0.558 | 2.660 | 0.830 | 0.058 | 6-311+g(d) |
| | 0.558 | -2.102 | -0.772 | 2.660 | 0.112 | 2.102 | -0.558 | 2.660 | 0.830 | 0.058 | 6-311++g(d) |
| $Be^-$ | -0.232 | -3.833 | -2.032 | 3.601 | 0.574 | 3.833 | 0.232 | 3.601 | 2.040 | 0.007 | 6-31+g(d) |
| | -0.227 | -17.951 | -9.089 | 17.724 | 2.330 | 17.951 | 0.227 | 17.724 | 9.090 | 0.001 | 6-311+g(d) |
| | -0.227 | -17.951 | -9.089 | 17.724 | 2.330 | 17.951 | 0.227 | 17.724 | 9.090 | 0.001 | 6-311++g(d) |
| $B^-$ | -0.326 | -3.819 | -2.073 | 3.494 | 0.615 | 3.819 | 0.326 | 3.494 | 2.088 | 0.015 | 6-31+g(d) |
| | -0.350 | -35.493 | -17.922 | 35.142 | 4.570 | 35.493 | 0.350 | 35.142 | 17.923 | 0.002 | 6-311+g(d) |
| | -0.350 | -35.493 | -17.922 | 35.142 | 4.570 | 35.493 | 0.350 | 35.142 | 17.923 | 0.002 | 6-311++g(d) |
| $C^-$ | 1.672 | -5.641 | -1.985 | 7.313 | 0.269 | 5.641 | -1.672 | 7.313 | 2.176 | 0.191 | 6-31+g(d) |
| | 1.642 | -5.716 | -2.037 | 7.358 | 0.282 | 5.716 | -1.642 | 7.358 | 2.220 | 0.183 | 6-311+g(d) |
| | 1.642 | -5.716 | -2.037 | 7.358 | 0.282 | 5.716 | -1.642 | 7.358 | 2.220 | 0.183 | 6-311++g(d) |
| $N^-$ | 1.072 | -5.156 | -2.042 | 6.228 | 0.335 | 5.156 | -1.072 | 6.228 | 2.134 | 0.092 | 6-31+g(d) |
| | 1.058 | -5.179 | -2.060 | 6.238 | 0.340 | 5.179 | -1.058 | 6.238 | 2.150 | 0.090 | 6-311+g(d) |
| | 1.058 | -5.179 | -2.060 | 6.238 | 0.340 | 5.179 | -1.058 | 6.238 | 2.150 | 0.090 | 6-311++g(d) |
| $O^-$ | 4.377 | -7.243 | -1.433 | 11.620 | 0.088 | 7.243 | -4.377 | 11.620 | 2.257 | 0.092 | 6-31+g(d) |
| | 4.357 | -7.265 | -1.454 | 11.621 | 0.091 | 7.265 | -4.357 | 11.621 | 2.271 | 0.817 | 6-311+g(d) |
| | 4.357 | -7.265 | -1.454 | 11.621 | 0.091 | 7.265 | -4.357 | 11.621 | 2.271 | 0.817 | 6-311++g(d) |
| $F^-$ | 3.513 | -13.183 | -4.835 | 16.696 | 0.700 | 13.183 | -3.513 | 16.696 | 5.205 | 0.369 | 6-31+g(d) |
| | 3.486 | -13.355 | -4.934 | 16.841 | 0.723 | 13.355 | -3.486 | 16.841 | 5.295 | 0.361 | 6-311+g(d) |
| | 3.486 | -13.355 | -4.934 | 16.841 | 0.723 | 13.355 | -3.486 | 16.841 | 5.295 | 0.361 | 6-311++g(d) |
| $Ne^-$ | -6.833 | -14.840 | -10.836 | 8.006 | 7.334 | 14.840 | 6.833 | 8.006 | 13.753 | 2.916 | 6-31+g(d) |
| | -6.584 | -14.472 | -10.528 | 7.888 | 7.025 | 14.472 | 6.584 | 7.888 | 13.275 | 2.747 | 6-311+g(d) |
| | -6.584 | -14.472 | -10.528 | 7.888 | 7.025 | 14.472 | 6.584 | 7.888 | 13.275 | 2.747 | 6-311++g(d) |

(e)

| Ions | I (eV) | E (eV) | χ (eV) | η (eV) | ω (eV) | $\mu^+$ (eV) | $\mu^-$ (eV) | η (eV) | $\omega^+$ (eV) | $\omega^-$ (eV) | Basis sets |
|---|---|---|---|---|---|---|---|---|---|---|---|
| $Li^{2-}$ | -2.190 | -3.891 | -3.041 | 1.700 | 2.718 | 3.891 | 2.190 | 1.700 | 4.452 | 1.411 | 6-31+g(d) |
| | -2.102 | -3.680 | -2.891 | 1.578 | 2.647 | 3.680 | 2.102 | 1.578 | 4.290 | 1.399 | 6-311+g(d) |
| | -2.102 | -3.680 | -2.891 | 1.578 | 2.647 | 3.680 | 2.102 | 1.578 | 4.290 | 1.399 | 6-311++g(d) |
| $Be^{2-}$ | -3.833- | -6.431 | -5.162 | 2.598 | 5.068 | 6.431 | 3.833 | 2.598 | 7.959 | 2.827 | 6-31+g(d) |
| | 17.950 | -31.972 | -24.961 | 14.022 | 22.218 | 31.972 | 17.950 | 14.022 | 36.452 | 11.490 | 6-311+g(d) |
| | -17.950 | -31.972 | -24.961 | 14.022 | 22.218 | 31.972 | 17.950 | 14.022 | 36.452 | 11.490 | 6-311++g(d) |
| $B^{2-}$ | -3.819- | -84.256 | -44.038 | 80.436 | 12.055 | 84.256 | 3.819 | 80.436 | 44.128 | 0.091 | 6-31+g(d) |
| | 35.493 | -46.585 | -41.039 | 11.093 | 75.916 | 46.585 | 35.493 | 11.093 | 97.822 | 56.783 | 6-311+g(d) |
| | -35.493 | -46.585 | -41.039 | 11.093 | 75.916 | 46.585 | 35.493 | 11.093 | 97.822 | 56.783 | 6-311++g(d) |
| $C^{2-}$ | -5.641- | -117.047 | -61.344 | 111.406 | 16.889 | 117.047 | 5.641 | 111.406 | 44.128 | 0.143 | 6-31+g(d) |
| | 5.716 | -110.654 | -58.185 | 104.938 | 16.131 | 110.654 | 5.716 | 104.938 | 58.341 | 0.156 | 6-311+g(d) |
| | -5.716 | -110.654 | -58.185 | 104.938 | 16.131 | 110.654 | 5.716 | 104.938 | 58.341 | 0.156 | 6-311++g(d) |
| $N^{2-}$ | -5.156 | -12.436 | -8.796 | 7.280 | 5.314 | 12.436 | 5.156 | 7.280 | 10.622 | 1.826 | 6-31+g(d) |
| | -5.179 | -12.386 | -8.783 | 7.207 | 5.352 | 12.386 | 5.179 | 7.207 | 10.644 | 1.861 | 6-311+g(d) |
| | -5.179 | -12.386 | -8.783 | 7.207 | 5.352 | 12.386 | 5.179 | 7.207 | 10.644 | 1.861 | 6-311++g(d) |
| $O^{2-}$ | -7.243 | -17.019 | -12.131 | 9.775 | 7.527 | 17.019 | 7.243 | 9.775 | 14.815 | 2.684 | 6-31+g(d) |
| | -7.265 | -17.278 | -12.271 | 10.013 | 7.512 | 17.278 | 7.265 | 10.013 | 14.907 | 2.635 | 6-311+g(d) |
| | -7.265 | -17.278 | -12.271 | 10.013 | 7.512 | 17.278 | 7.265 | 10.013 | 14.907 | 2.635 | 6-311++g(d) |
| $F^{2-}$ | -13.183- | -19.975 | -16.579 | 6.793 | 20.233 | 19.975 | 13.183 | 6.793 | 29.372 | 12.793 | 6-31+g(d) |
| | 13.355 | -20.248 | -16.802 | 6.893 | 20.476 | 20.248 | 13.355 | 6.893 | 29.738 | 12.936 | 6-311+g(d) |
| | -13.355 | -20.248 | -16.802 | 6.893 | 20.476 | 20.248 | 13.355 | 6.893 | 29.738 | 12.936 | 6-311++g(d) |
| $Ne^{2-}$ | -14.839 | -22.975 | -18.907 | 8.135 | 21.972 | 22.975 | 14.839 | 8.135 | 332.442 | 13.535 | 6-31+g(d) |
| | -14.472 | -22.368 | -18.420 | 7.896 | 21.485 | 22.368 | 14.472 | 7.896 | 31.682 | 13.261 | 6-311+g(d) |
| | -14.472 | -22.368 | -18.420 | 7.896 | 21.485 | 22.368 | 14.472 | 7.896 | 31.682 | 13.261 | 6-311++g(d) |

**Table 6**: Ionization potential(I) , Electron affinity(E), Electronegativity(χ), Chemical hardness(η) , Electrophilicity(ω) and the values of ($\mu^+$, $\mu^-$ ,$\omega^+$ ,$\omega^-$ ) for Electroaccepting and Electrodonating processes of atoms and ions by using Hartree-Fock method in aqueous phase

(a)

| **Atoms** | **I (eV)** | **E (eV)** | **χ (eV)** | **η (eV)** | **ω (eV)** | **$\mu^+$ (eV)** | **$\mu^-$ (eV)** | **η (eV)** | **$\omega^+$ (eV)** | **$\omega^-$ (eV)** | **Basis sets** |
|---|---|---|---|---|---|---|---|---|---|---|---|
| Li | 0.866 | -0.013 | 0.427 | 0.878 | 0.104 | 0.013 | -0.866 | 0.879 | 0.000 | 0.427 | 6-31+g(d) |
| | 0.898 | -0.016 | 0.441 | 0.914 | 0.106 | 0.016 | -0.898 | 0.914 | 0.000 | 0.441 | 6-311+g(d) |
| | 0.898 | -0.016 | 0.441 | 0.915 | 0.106 | 0.016 | -0.898 | 0.914 | 0.000 | 0.441 | 6-311++g(d) |
| Be | 3.513 | 0.866 | 2.190 | 2.647 | 0.906 | -0.866 | -3.513 | 2.647 | 0.142 | 2.331 | 6-31+g(d) |
| | 3.562 | 0.943 | 2.254 | 2.619 | 0.968 | -0.943 | -3.562 | 2.619 | 0.170 | 2.422 | 6-311+g(d) |
| | 3.562 | 0.943 | 2.254 | 2.619 | 0.968 | -0.943 | -3.562 | 2.619 | 0.170 | 2.422 | 6-311++g(d) |
| B | 5.098 | 0.931 | 3.014 | 4.167 | 1.090 | -0.931 | -5.098 | 4.167 | 0.104 | 3.119 | 6-31+g(d) |
| | 5.173 | 1.019 | 3.096 | 4.153 | 1.154 | -1.019 | -5.173 | 4.154 | 0.125 | 3.221 | 6-311+g(d) |
| | 5.173 | 1.019 | 3.096 | 4.153 | 1.154 | -1.019 | -5.173 | 4.154 | 0.125 | 3.221 | 6-311++g(d) |
| C | 5.376 | 3.386 | 4.382 | 1.989 | 4.825 | -3.386 | -5.376 | 1.990 | 2.883 | 7.264 | 6-31+g(d) |
| | 5.400 | 3.440 | 4.420 | 1.960 | 4.984 | -3.440 | -5.400 | 1.960 | 3.018 | 7.438 | 6-311+g(d) |
| | 5.400 | 3.440 | 4.420 | 1.960 | 4.984 | -3.440 | -5.400 | 1.960 | 3.018 | 7.438 | 6-311++g(d) |
| N | 10.423 | 2.384 | 6.404 | 8.038 | 2.551 | -2.384 | -10.423 | 8.039 | 0.354 | 6.757 | 6-31+g(d) |
| | 10.388 | 2.378 | 6.383 | 7.820 | 2.528 | -2.378 | -10.388 | 7.820 | 0.362 | 6.649 | 6-311+g(d) |
| | 10.388 | 2.378 | 6.383 | 8.010 | 2.528 | -2.378 | -10.388 | 8.010 | 0.353 | 6.736 | 6-311++g(d) |
| O | 10.281 | 6.154 | 8.217 | 4.127 | 8.181 | -6.154 | -10.281 | 4.127 | 4.588 | 12.806 | 6-31+g(d) |
| | 10.240 | 6.136 | 8.188 | 4.104 | 8.167 | -6.136 | -10.240 | 4.104 | 4.586 | 12.774 | 6-311+g(d) |
| | 10.240 | 6.136 | 8.188 | 4.104 | 8.167 | -6.136 | -10.240 | 4.104 | 4.586 | 12.774 | 6-311++g(d) |
| F | 16.607 | 4.708 | 10.657 | 11.899 | 4.773 | -4.708 | -16.607 | 11.899 | 0.931 | 11.589 | 6-31+g(d) |
| | 16.593 | 4.384 | 10.638 | 11.909 | 4.752 | -4.384 | -16.593 | 11.909 | 0.921 | 11.600 | 6-311+g(d) |
| | 16.593 | 4.384 | 10.638 | 11.909 | 4.752 | -4.384 | -16.593 | 11.909 | 0.921 | 11.600 | 6-311++g(d) |
| Ne | 16.267 | -4.359 | 5.954 | 20.626 | 0.893 | 4.359 | -16.267 | 20.626 | 0.461 | 6.414 | 6-31+g(d) |
| | 16.170 | -4.068 | 6.051 | 20.238 | 0.904 | 4.068 | -16.170 | 20.238 | 0.409 | 6.460 | 6-311+g(d) |
| | 16.170 | -4.068 | 6.051 | 20.237 | 0.904 | 4.068 | -16.170 | 20.238 | 0.408 | 6.450 | 6-311++g(d) |

(b)

| **Ions** | **I (eV)** | **E (eV)** | **χ (eV)** | **η (eV)** | **ω (eV)** | **$\mu^+$ (eV)** | **$\mu^-$ (eV)** | **η (eV)** | **$\omega^+$ (eV)** | **$\omega^-$ (eV)** | **Basis sets** |
|---|---|---|---|---|---|---|---|---|---|---|---|
| $Li^+$ | 61.262 | 0.866 | 31.064 | 60.396 | 7.988 | -0.866 | -61.262 | 60.396 | 0.006 | 31.070 | 6-31+g(d) |
| | 60.239 | 0.898 | 30.569 | 59.341 | 7.874 | -0.898 | -60.239 | 59.341 | 0.007 | 30.576 | 6-311+g(d) |
| | 60.239 | 0.898 | 30.569 | 59.341 | 7.874 | -0.898 | -60.239 | 59.341 | 0.007 | 30.576 | 6-311++g(d) |
| $Be^+$ | 5.563 | 3.613 | 4.538 | 2.050 | 5.024 | -3.613 | -5.563 | 2.050 | 3.011 | 7.549 | 6-31+g(d) |
| | 5.561 | 3.562 | 4.561 | 1.999 | 5.204 | -3.562 | -5.561 | 1.999 | 3.173 | 7.734 | 6-311+g(d) |
| | 5.561 | 3.562 | 4.561 | 1.999 | 5.204 | -3.562 | -5.561 | 1.999 | 3.173 | 7.734 | 6-311++g(d) |
| $B^+$ | 14.827 | 5.098 | 9.962 | 9.728 | 5.101 | -5.098 | -14.827 | 9.729 | 1.336 | 11.298 | 6-31+g(d) |
| | 14.819 | 5.173 | 9.996 | 9.646 | 5.179 | -5.173 | -14.819 | 9.646 | 1.387 | 11.383 | 6-311+g(d) |
| | 14.819 | 5.173 | 9.996 | 9.646 | 5.179 | -5.173 | -14.819 | 9.646 | 1.387 | 11.383 | 6-311++g(d) |
| $C^+$ | 14.999 | 5.376 | 10.188 | 9.623 | 5.393 | -5.376 | -14.999 | 9.623 | 1.502 | 11.689 | 6-31+g(d) |
| | 15.010 | 5.400 | 10.205 | 9.610 | 5.418 | -5.400 | -15.010 | 9.610 | 1.517 | 11.722 | 6-311+g(d) |
| | 15.010 | 5.400 | 10.205 | 9.610 | 5.418 | -5.400 | -15.010 | 9.610 | 1.517 | 11.722 | 6-311++g(d) |
| $N^+$ | 16.244 | 10.423 | 13.333 | 5.821 | 15.271 | -10.423 | -16.244 | 5.821 | 9.332 | 22.665 | 6-31+g(d) |
| | 16.186 | 10.388 | 13.287 | 6.178 | 14.288 | -10.388 | -16.186 | 6.178 | 8.417 | 21.722 | 6-311+g(d) |
| | 16.186 | 10.388 | 13.287 | 5.798 | 14.288 | -10.388 | -16.186 | 5.798 | 9.306 | 22.592 | 6-311++g(d) |
| $O^+$ | 23.864 | 10.281 | 17.072 | 13.583 | 10.729 | -10.281 | -23.864 | 13.583 | 3.891 | 20.963 | 6-31+g(d) |
| | 23.817 | 10.240 | 17.028 | 13.577 | 10.678 | -10.240 | -23.817 | 13.577 | 3.861 | 21.704 | 6-311+g(d) |
| | 23.817 | 10.240 | 17.028 | 13.577 | 10.678 | -10.240 | -23.817 | 13.577 | 3.861 | 20.890 | 6-311++g(d) |
| $F^+$ | 24.799 | 16.607 | 20.703 | 8.192 | 26.161 | -16.607 | -24.799 | 8.192 | 16.833 | 37.536 | 6-31+g(d) |
| | 24.722 | 16.593 | 20.657 | 8.129 | 26.248 | -16.593 | -24.722 | 8.129 | 16.935 | 37.593 | 6-311+g(d) |
| | 24.722 | 16.593 | 20.657 | 8.129 | 26.248 | -16.593 | -24.722 | 8.129 | 16.935 | 37.593 | 6-311++g(d) |
| $Ne^+$ | 33.724 | 16.267 | 24.996 | 17.457 | 17.894 | -16.267 | -33.724 | 17.457 | 7.579 | 32.574 | 6-31+g(d) |
| | 33.711 | 16.170 | 24.940 | 17.541 | 17.730 | -16.170 | -33.711 | 17.541 | 7.453 | 32.393 | 6-311+g(d) |
| | 33.711 | 16.170 | 24.940 | 17.541 | 17.730 | -16.170 | -33.711 | 17.541 | 7.453 | 32.393 | 6-311++g(d) |

(c)

| Ions | I (eV) | E (eV) | χ (eV) | η (eV) | ω (eV) | $\mu^+$ (eV) | $\mu^-$ (eV) | η (eV) | $\omega^+$ (eV) | $\omega^-$ (eV) | Basis sets |
|---|---|---|---|---|---|---|---|---|---|---|---|
| $Li^{2+}$ | 97.307 | 61.261 | 79.285 | 36.045 | 87.196 | -61.261 | -97.307 | 36.046 | 52.059 | 131.344 | 6-31+g(d) |
| | 98.338 | 60.239 | 79.288 | 38.098 | 82.506 | -60.239 | -98.338 | 38.099 | 47.624 | 126.912 | 6-311+g(d) |
| | 98.338 | 60.239 | 79.288 | 38.096 | 82.506 | -60.239 | -98.338 | 38.099 | 47.624 | 126.912 | 6-311++g(d) |
| $Be^{2+}$ | 131.864 | 5.563 | 68.713 | 126.301 | 18.691 | -5.563 | -131.864 | 126.301 | 0.123 | 68.836 | 6-31+g(d) |
| | 131.493 | 5.561 | 68.527 | 125.932 | 18.645 | -5.561 | -131.493 | 125.932 | 0.123 | 68.649 | 6-311+g(d) |
| | 131.493 | 5.561 | 68.527 | 125.932 | 18.645 | -5.561 | -131.493 | 125.932 | 0.123 | 68.650 | 6-311++g(d) |
| $B^{2+}$ | 23.404 | 14.827 | 19.115 | 8.577 | 21.300 | -14.827 | -23.404 | 8.577 | 12.814 | 31.930 | 6-31+g(d) |
| | 23.425 | 14.819 | 19.122 | 8.606 | 21.243 | -14.819 | -23.425 | 8.606 | 12.758 | 31.879 | 6-311+g(d) |
| | 23.425 | 14.819 | 19.122 | 8.606 | 21.243 | -14.819 | -23.425 | 8.606 | 12.758 | 31.880 | 6-311++g(d) |
| $C^{2+}$ | 30.515 | 14.999 | 22.757 | 15.516 | 16.689 | -14.999 | -30.515 | 15.516 | 7.250 | 30.007 | 6-31+g(d) |
| | 30.555 | 15.010 | 22.783 | 15.546 | 16.694 | -15.010 | -30.555 | 15.545 | 7.246 | 30.029 | 6-311+g(d) |
| | 30.555 | 15.010 | 22.783 | 15.546 | 16.694 | -15.010 | -30.555 | 15.545 | 7.246 | 30.029 | 6-311++g(d) |
| $N^{2+}$ | 31.236 | 16.244 | 23.740 | 14.992 | 18.796 | -16.244 | -31.236 | 14.992 | 8.800 | 32.540 | 6-31+g(d) |
| | 31.276 | 16.186 | 23.731 | 14.900 | 19.048 | -16.186 | -31.276 | 14.900 | 8.998 | 32.824 | 6-311+g(d) |
| | 31.276 | 16.186 | 23.731 | 15.090 | 19.048 | -16.186 | -31.276 | 15.090 | 8.680 | 32.411 | 6-311++g(d) |
| O2+ | 33.617 | 23.864 | 28.741 | 9.753 | 42.346 | -23.864 | -33.617 | 9.753 | 29.195 | 57.936 | 6-31+g(d) |
| | 33.517 | 23.817 | 28.667 | 9.699 | 42.362 | -23.817 | -33.517 | 9.700 | 29.241 | 57.908 | 6-311+g(d) |
| | 33.517 | 23.817 | 28.667 | 9.700 | 42.362 | -23.817 | -33.517 | 9.700 | 29.241 | 57.908 | 6-311++g(d) |
| $F^{2+}$ | 43.791 | 24.799 | 34.295 | 18.992 | 30.965 | -24.799 | -43.791 | 18.992 | 16.192 | 50.487 | 6-31+g(d) |
| | 43.770 | 24.722 | 34.246 | 19.047 | 30.786 | -24.722 | -43.770 | 19.048 | 16.044 | 50.290 | 6-311+g(d) |
| | 43.770 | 24.722 | 34.246 | 19.047 | 30.786 | -24.722 | -43.770 | 19.048 | 16.044 | 50.290 | 6-311++g(d) |
| $Ne^{2+}$ | 45.837 | 33.724 | 39.781 | 12.113 | 65.323 | -33.724 | -45.837 | 12.113 | 46.947 | 86.728 | 6-31+g(d) |
| | 45.766 | 33.711 | 39.738 | 12.056 | 65.494 | -33.711 | -45.766 | 12.055 | 47.132 | 86.871 | 6-311+g(d) |
| | 45.766 | 33.711 | 39.738 | 12.056 | 65.494 | -33.711 | -45.766 | 12.055 | 47.132 | 86.871 | 6-311++g(d) |

(d)

| Ions | I (eV) | E (eV) | χ (eV) | η (eV) | ω (eV) | $\mu^+$ (eV) | $\mu^-$ (eV) | η (eV) | $\omega^+$ (eV) | $\omega^-$ (eV) | Basis sets |
|---|---|---|---|---|---|---|---|---|---|---|---|
| $Li^-$ | -0.013 | -1.278 | -0.646 | 1.266 | 0.165 | 1.278 | 0.013 | 1.265 | 0.646 | 0.000 | 6-31+g(d) |
| | -0.016 | -1.241 | -0.628 | 1.224 | 0.161 | 1.241 | 0.016 | 1.225 | 0.628 | 0.000 | 6-311+g(d) |
| | -0.016 | -1.241 | -0.628 | 1.224 | 0.161 | 1.241 | 0.016 | 1.225 | 0.628 | 0.000 | 6-311++g(d) |
| $Be^-$ | 0.866 | -0.239 | 0.314 | 1.105 | 0.045 | 0.239 | -0.866 | 1.105 | 0.026 | 0.339 | 6-31+g(d) |
| | 0.943 | -0.234 | 0.354 | 1.176 | 0.053 | 0.234 | -0.943 | 1.177 | 0.023 | 0.377 | 6-311+g(d) |
| | 0.943 | -0.234 | 0.354 | 1.176 | 0.053 | 0.234 | -0.943 | 1.177 | 0.023 | 0.378 | 6-311++g(d) |
| $B^-$ | 0.931 | 1.301 | 1.116 | -0.370 | -1.683 | -1.301 | -0.931 | -0.370 | -2.288 | -1.171 | 6-31+g(d) |
| | 1.019 | 1.358 | 1.189 | -0.338 | -2.088 | -1.358 | -1.019 | -0.339 | -2.725 | -1.536 | 6-311+g(d) |
| | 1.019 | 1.358 | 1.189 | -0.338 | -2.088 | -1.358 | -1.019 | -0.339 | -2.725 | -1.536 | 6-311++g(d) |
| $C^-$ | 3.387 | 0.153 | 1.770 | 3.234 | 0.484 | -0.153 | -3.387 | 3.234 | 0.004 | 1.773 | 6-31+g(d) |
| | 3.440 | 0.169 | 1.804 | 3.271 | 0.498 | -0.169 | -3.440 | 3.271 | 0.004 | 1.809 | 6-311+g(d) |
| | 3.440 | 0.169 | 1.804 | 3.271 | 0.498 | -0.169 | -3.440 | 3.271 | 0.004 | 1.809 | 6-311++g(d) |
| $O^-$ | 6.154 | -0.200 | 2.797 | 6.354 | 0.697 | 0.200 | -6.154 | 6.354 | 0.003 | 2.980 | 6-31+g(d) |
| | 6.136 | -0.206 | 2.965 | 6.342 | 0.693 | 0.206 | -6.136 | 6.342 | 0.003 | 2.968 | 6-311+g(d) |
| | 6.136 | -0.206 | 2.965 | 6.342 | 0.693 | 0.206 | -6.136 | 6.342 | 0.003 | 2.968 | 6-311++g(d) |
| $F^-$ | 4.708 | -4.351 | 0.178 | 9.058 | 0.002 | 4.351 | -4.708 | 9.059 | 1.044 | 1.223 | 6-31+g(d) |
| | 4.684 | -4.459 | 0.112 | 9.143 | 0.001 | 4.459 | -4.684 | 9.143 | 1.088 | 1.199 | 6-311+g(d) |
| | 4.684 | -4.459 | 0.112 | 9.143 | 0.001 | 4.459 | -4.684 | 9.143 | 1.087 | 1.199 | 6-311++g(d) |
| $Ne^-$ | -4.359 | -5.848 | -5.104 | 1.488 | 8.750 | 5.848 | 4.359 | 1.489 | 11.488 | 6.385 | 6-31+g(d) |
| | -4.068 | -5.452 | -4.760 | 1.384 | 8.184 | 5.452 | 4.068 | 1.384 | 10.736 | 5.977 | 6-311+g(d) |
| | -4.068 | -5.452 | -4.760 | 1.384 | 8.184 | 5.452 | 4.068 | 1.384 | 10.736 | 5.977 | 6-311++g(d) |

(e)

| Ions | I (eV) | E (eV) | χ (eV) | η (eV) | ω (eV) | $\mu^+$ (eV) | $\mu^-$ (eV) | η (eV) | $\omega^+$ (eV) | $\omega^-$ (eV) | Basis sets |
|---|---|---|---|---|---|---|---|---|---|---|---|
| $Li^{2-}$ | -1.278 | -2.393 | -1.836 | 1.114 | 1.512 | 2.393 | 1.278 | 1.115 | 2.569 | 0.734 | 6-31+g(d) |
| | -1.241 | -2.371 | -1.806 | 1.131 | 1.442 | 2.371 | 1.241 | 1.130 | 2.487 | 0.681 | 6-311+g(d) |
| | -1.241 | -2.371 | -1.806 | 1.131 | 1.442 | 2.371 | 1.241 | 1.130 | 2.487 | 0.681 | 6-311++g(d) |
| $Be^{2-}$ | -0.239 | -0.103 | -0.171 | -0.136 | -0.107 | 0.103 | 0.239 | -0.137 | -0.039 | -0.209 | 6-31+g(d) |
| | -0.239 | -0.103 | -0.701 | -0.128 | -0.112 | 0.103 | 0.239 | -0.126 | -0.044 | -0.213 | 6-311+g(d) |
| | -0.234 | -0.106 | -0.170 | -0.128 | -0.112 | 0.106 | 0.234 | -0.128 | -0.044 | -0.213 | 6-311++g(d) |
| $B^{2-}$ | 1.301 | -0.040 | 0.631 | 1.341 | 0.148 | 0.040 | -1.301 | 1.341 | 0.001 | 0.631 | 6-31+g(d) |
| | 1.358 | 0.030 | 0.694 | 1.327 | 0.181 | -0.030 | -1.358 | 1.328 | 0.000 | 0.694 | 6-311+g(d) |
| | 1.358 | 0.030 | 0.694 | 1.327 | 0.181 | -0.030 | -1.358 | 1.328 | 0.000 | 0.694 | 6-311++g(d) |
| $C^{2-}$ | 0.153 | 0.871 | 0.512 | -0.719 | -0.182 | -0.871 | -0.153 | -0.718 | -0.528 | -0.016 | 6-31+g(d) |
| | 0.169 | 0.881 | 0.525 | -0.712 | -0.194 | -0.881 | -0.169 | -0.712 | -0.545 | -0.020 | 6-311+g(d) |
| | 0.169 | 0.881 | 0.525 | -0.712 | -0.194 | -0.881 | -0.169 | -0.712 | -0.545 | -0.020 | 6-311++g(d) |
| $O^{2-}$ | -0.200 | -2.451 | -1.326 | 2.251 | 0.390 | 2.451 | 0.200 | 2.251 | 1.335 | 0.009 | 6-31+g(d) |
| | -0.206 | -2.607 | -1.406 | 2.400 | 0.412 | 2.607 | 0.206 | 2.401 | 1.415 | 0.009 | 6-311+g(d) |
| | -0.206 | -2.607 | -1.406 | 2.400 | 0.412 | 2.607 | 0.206 | 2.401 | 1.415 | 0.009 | 6-311++g(d) |
| $F^{2-}$ | -4.351 | -5.134 | -4.742 | 0.783 | 14.366 | 5.134 | 4.351 | 0.783 | 16.835 | 12.093 | 6-31+g(d) |
| | -4.459 | -5.316 | -4.888 | 0.857 | 13.939 | 5.316 | 4.459 | 0.857 | 16.499 | 11.602 | 6-311+g(d) |
| | -4.459 | -5.316 | -4.888 | 0.857 | 13.939 | 5.316 | 4.459 | 0.857 | 16.499 | 11.602 | 6-311++g(d) |
| $Ne^{2-}$ | -5.847 | -6.485 | -6.166 | 0.638 | 29.811 | 6.485 | 5.847 | 0.638 | 32.974 | 26.807 | 6-31+g(d) |
| | -5.452 | -5.976 | -5.714 | 0.524 | 31.158 | 5.976 | 5.452 | 0.524 | 34.080 | 28.367 | 6-311+g(d) |
| | -5.452 | -5.976 | -5.714 | 0.524 | 31.158 | 5.976 | 5.452 | 0.524 | 34.080 | 28.367 | 6-311++g(d) |

**Table 7**: Ionization potential(I) , Electron affinity(E), Electronegativity(χ), Chemical hardness(η) , Electrophilicity(ω) and the values of ($\mu^+$, $\mu^-$ ,$\omega^+$ ,$\omega^-$ ) for Electroaccepting and Electrodonating processes of atoms and ions by using MP2 method in aqueous phase

(a)

| Atoms | I (eV) | E (eV) | χ (eV) | η (eV) | ω (eV) | $\mu^+$ (eV) | $\mu^-$ (eV) | η (eV) | $\omega^+$ (eV) | $\omega^-$ (eV) | Basis sets |
|---|---|---|---|---|---|---|---|---|---|---|---|
| Li | 0.866 | -0.013 | 0.427 | 0.878 | 0.104 | 0.013 | -0.866 | 0.879 | 0.000 | 0.427 | 6-31+g(d) |
| | 0.898 | -0.016 | 0.441 | 0.915 | 0.106 | 0.016 | -0.898 | 0.914 | 0.000 | 0.441 | 6-311+g(d) |
| | 0.898 | -0.016 | 0.451 | 0.915 | 0.106 | 0.016 | -0.898 | 0.914 | 0.000 | 0.441 | 6-311++g(d) |
| Be | 3.513 | 0.866 | 2.190 | 2.647 | 0.906 | -0.866 | -3.513 | 2.647 | 0.142 | 2.331 | 6-31+g(d) |
| | 3.562 | 0.943 | 2.525 | 2.619 | 0.968 | -0.943 | -3.562 | 2.619 | 0.170 | 2.422 | 6-311+g(d) |
| | 3.562 | 0.943 | 2.252 | 2.619 | 0.968 | -0.943 | -3.562 | 2.619 | 0.891 | 3.652 | 6-311++g(d) |
| B | 5.098 | 0.931 | 3.015 | 4.167 | 1.091 | -0.931 | -5.098 | 4.167 | 0.104 | 3.119 | 6-31+g(d) |
| | 5.173 | 1.019 | 3.096 | 4.153 | 1.154 | -1.019 | -5.173 | 4.154 | 0.125 | 3.221 | 6-311+g(d) |
| | 5.173 | 1.019 | 3.096 | 4.153 | 1.154 | -1.019 | -5.173 | 4.154 | 0.125 | 3.221 | 6-311++g(d) |
| C | 5.376 | 3.387 | 4.382 | 1.989 | 4.825 | -3.387 | -5.376 | 1.989 | 2.883 | 7.264 | 6-31+g(d) |
| | 5.400 | 3.440 | 4.420 | 1.960 | 4.983 | -3.440 | -5.400 | 1.960 | 3.018 | 7.438 | 6-311+g(d) |
| | 5.330 | 3.440 | 4.420 | 1.891 | 5.085 | -3.440 | -5.330 | 1.891 | 3.129 | 7.514 | 6-311++g(d) |
| N | 10.423 | 2.384 | 6.404 | 8.038 | 2.551 | -2.384 | -10.423 | 8.039 | 0.354 | 6.757 | 6-31+g(d) |
| | 10.388 | 2.378 | 6.383 | 8.010 | 2.543 | -2.378 | -10.388 | 8.010 | 0.353 | 6.736 | 6-311+g(d) |
| | 8.840 | 2.378 | 6.383 | 8.010 | 2.543 | -2.378 | -8.840 | 8.010 | 0.353 | 6.736 | 6-311++g(d) |
| O | 10.281 | 6.154 | 8.217 | 4.127 | 8.181 | -6.154 | -10.281 | 4.127 | 4.588 | 12.806 | 6-31+g(d) |
| | 10.240 | 6.136 | 8.188 | 4.104 | 8.167 | -6.136 | -10.240 | 4.104 | 4.586 | 12.774 | 6-311+g(d) |
| | 10.388 | 6.136 | 8.188 | 4.104 | 8.167 | -6.136 | -10.388 | 4.104 | 4.586 | 12.774 | 6-311++g(d) |
| F | 16.607 | 4.708 | 10.657 | 11.899 | 4.773 | -4.708 | -16.607 | 11.899 | 0.931 | 11.589 | 6-31+g(d) |
| | 16.593 | 4.684 | 10.638 | 11.909 | 4.752 | -4.684 | -16.593 | 11.909 | 0.921 | 11.600 | 6-311+g(d) |
| | 16.593 | 4.684 | 10.638 | 11.909 | 4.752 | -4.684 | -16.593 | 11.909 | 0.921 | 11.559 | 6-311++g(d) |
| Ne | 16.167 | -4.359 | 5.954 | 20.626 | 0.859 | 4.359 | -16.167 | 20.626 | 0.461 | 6.414 | 6-31+g(d) |
| | 16.170 | -4.068 | 6.051 | 20.237 | 0.905 | 4.068 | -16.170 | 20.238 | 0.409 | 6.500 | 6-311+g(d) |
| | 16.170 | -4.068 | 6.051 | 20.237 | 0.905 | 4.068 | -16.170 | 20.238 | 0.408 | 6.500 | 6-311++g(d) |

(b)

| Ions | I (eV) | E (eV) | $\chi$ (eV) | $\eta$ (eV) | $\omega$ (eV) | $\mu^+$ (eV) | $\mu^-$ (eV) | $\eta$ (eV) | $\omega^+$ (eV) | $\omega^-$ (eV) | Basis sets |
|---|---|---|---|---|---|---|---|---|---|---|---|
| $Li^+$ | 61.262 | 0.866 | 31.064 | 60.396 | 7.988 | -0.866 | -61.262 | 60.396 | 0.006 | 31.070 | 6-31+g(d) |
| | 60.239 | 0.898 | 30.569 | 59.341 | 7.874 | -0.898 | -60.239 | 59.341 | 0.007 | 30.576 | 6-311+g(d) |
| | 60.239 | 0.898 | 30.569 | 59.341 | 7.874 | -0.898 | -60.239 | 59.341 | 0.007 | 30.576 | 6-311++g(d) |
| $Be^+$ | 5.563 | 3.513 | 4.538 | 2.050 | 5.024 | -3.513 | -5.563 | 2.050 | 3.011 | 7.549 | 6-31+g(d) |
| | 5.561 | 3.562 | 4.561 | 1.999 | 5.204 | -3.562 | -5.561 | 1.999 | 3.173 | 7.734 | 6-311+g(d) |
| | 5.561 | 3.562 | 4.561 | 1.999 | 5.203 | -3.562 | -5.561 | 1.999 | 3.173 | 7.734 | 6-311++g(d) |
| $B^+$ | 14.827 | 5.098 | 9.962 | 9.728 | 5.101 | -5.098 | -14.827 | 9.729 | 1.336 | 11.298 | 6-31+g(d) |
| | 14.819 | 5.173 | 9.996 | 9.646 | 5.179 | -5.173 | -14.819 | 9.646 | 1.387 | 11.383 | 6-311+g(d) |
| | 14.819 | 5.173 | 9.996 | 9.646 | 5.179 | -5.173 | -14.819 | 9.646 | 1.387 | 11.383 | 6-311++g(d) |
| $C^+$ | 14.999 | 5.376 | 10.188 | 9.623 | 5.393 | -5.376 | -14.999 | 9.623 | 1.502 | 11.690 | 6-31+g(d) |
| | 15.010 | 5.400 | 10.205 | 9.610 | 5.418 | -5.400 | -15.010 | 9.610 | 1.517 | 11.722 | 6-311+g(d) |
| | 15.010 | 5.331 | 10.205 | 9.748 | 5.341 | -5.331 | -15.010 | 9.749 | 1.457 | 11.663 | 6-311++g(d) |
| $N^+$ | 16.244 | 10.423 | 13.333 | 5.821 | 15.729 | -10.423 | -16.244 | 5.821 | 9.332 | 22.665 | 6-31+g(d) |
| | 16.186 | 10.380 | 13.287 | 5.798 | 15.224 | -10.380 | -16.186 | 5.806 | 9.305 | 22.592 | 6-311+g(d) |
| | 16.186 | 10.388 | 13.287 | 5.798 | 15.224 | -10.388 | -16.186 | 5.798 | 9.306 | 22.592 | 6-311++g(d) |
| $O^+$ | 23.864 | 10.281 | 17.072 | 13.583 | 10.729 | -10.281 | -23.864 | 13.583 | 3.891 | 20.963 | 6-31+g(d) |
| | 23.817 | 10.240 | 17.028 | 13.577 | 10.678 | -10.240 | -23.817 | 13.577 | 3.861 | 20.890 | 6-311+g(d) |
| | 23.817 | 10.240 | 17.028 | 13.577 | 10.678 | -10.240 | -23.817 | 13.577 | 3.861 | 20.890 | 6-311++g(d) |
| $F^+$ | 24.799 | 16.607 | 20.703 | 8.192 | 26.161 | -16.607 | -24.799 | 8.192 | 16.833 | 37.536 | 6-31+g(d) |
| | 24.722 | 16.593 | 20.658 | 8.129 | 26.248 | -16.593 | -24.722 | 8.129 | 16.935 | 37.593 | 6-311+g(d) |
| | 24.722 | 16.593 | 20.658 | 8.129 | 26.248 | -16.593 | -24.722 | 8.129 | 16.935 | 37.593 | 6-311++g(d) |
| $Ne^+$ | 33.724 | 16.267 | 24.996 | 17.457 | 17.894 | -16.267 | -33.724 | 17.457 | 7.578 | 32.574 | 6-31+g(d) |
| | 33.711 | 16.170 | 24.940 | 17.541 | 17.730 | -16.170 | -33.711 | 17.541 | 7.453 | 32.393 | 6-311+g(d) |
| | 33.711 | 16.170 | 24.940 | 17.541 | 17.730 | -16.170 | -33.711 | 17.541 | 7.453 | 32.393 | 6-311++g(d) |

(c)

| Ions | I (eV) | E (eV) | $\chi$ (eV) | $\eta$ (eV) | $\omega$ (eV) | $\mu^+$ (eV) | $\mu^-$ (eV) | $\eta$ (eV) | $\omega^+$ (eV) | $\omega^-$ (eV) | Basis sets |
|---|---|---|---|---|---|---|---|---|---|---|---|
| $Li^{2+}$ | 140.417 | 61.262 | 100.839 | 79.156 | 64.232 | -61.262 | -140.417 | 79.156 | 23.706 | 124.546 | 6-31+g(d) |
| | 141.448 | 60.239 | 100.844 | 81.208 | 62.613 | -60.239 | -141.448 | 81.209 | 22.342 | 123.186 | 6-311+g(d) |
| | 141.448 | 60.239 | 100.844 | 81.208 | 62.613 | -60.239 | -141.448 | 81.209 | 22.342 | 123.186 | 6-311++g(d) |
| $Be^{2+}$ | 131.864 | 5.563 | 68.713 | 126.301 | 18.691 | -5.563 | -131.864 | 126.301 | 0.123 | 68.836 | 6-31+g(d) |
| | 131.493 | 5.561 | 68.527 | 125.932 | 18.645 | -5.561 | -131.493 | 125.932 | 0.123 | 68.650 | 6-311+g(d) |
| | 131.493 | 5.561 | 68.527 | 125.932 | 18.645 | -5.561 | -131.493 | 125.932 | 0.123 | 68.650 | 6-311++g(d) |
| $B^{2+}$ | 23.404 | 14.827 | 19.115 | 8.577 | 21.300 | -14.827 | -23.404 | 8.577 | 12.814 | 31.930 | 6-31+g(d) |
| | 23.425 | 14.818 | 19.122 | 8.607 | 21.242 | -14.818 | -23.425 | 8.607 | 12.758 | 31.879 | 6-311+g(d) |
| | 23.425 | 14.819 | 19.122 | 8.607 | 21.243 | -14.819 | -23.425 | 8.607 | 12.758 | 31.879 | 6-311++g(d) |
| $C^{2+}$ | 30.515 | 14.999 | 22.757 | 15.516 | 16.689 | -14.999 | -30.515 | 15.516 | 7.250 | 30.007 | 6-31+g(d) |
| | 30.555 | 15.010 | 22.783 | 15.546 | 16.694 | -15.010 | -30.555 | 15.546 | 7.246 | 30.029 | 6-311+g(d) |
| | 30.555 | 15.010 | 22.783 | 15.546 | 16.694 | -15.010 | -30.555 | 15.546 | 7.246 | 30.029 | 6-311++g(d) |
| $N^{2+}$ | 31.236 | 16.244 | 23.740 | 14.992 | 18.796 | -16.244 | -31.236 | 14.992 | 8.800 | 32.540 | 6-31+g(d) |
| | 31.276 | 16.186 | 23.731 | 15.090 | 18.659 | -16.186 | -31.276 | 15.090 | 8.680 | 32.411 | 6-311+g(d) |
| | 31.276 | 16.186 | 23.731 | 15.090 | 18.659 | -16.186 | -31.276 | 15.090 | 8.680 | 32.411 | 6-311++g(d) |
| O2+ | 33.617 | 23.864 | 28.741 | 9.753 | 42.346 | -23.864 | -33.617 | 9.753 | 29.195 | 57.936 | 6-31+g(d) |
| | 33.517 | 23.817 | 28.667 | 9.699 | 42.362 | -23.817 | -33.517 | 9.699 | 29.241 | 57.908 | 6-311+g(d) |
| | 33.517 | 23.817 | 28.667 | 9.699 | 42.362 | -23.817 | -33.517 | 9.699 | 29.241 | 57.908 | 6-311++g(d) |
| $F^{2+}$ | 43.791 | 24.799 | 34.295 | 18.992 | 30.965 | -24.799 | -43.791 | 18.992 | 16.192 | 50.290 | 6-31+g(d) |
| | 43.769 | 24.722 | 34.246 | 19.047 | 30.786 | -24.722 | -43.769 | 19.047 | 16.044 | 50.290 | 6-311+g(d) |
| | 43.769 | 24.722 | 34.246 | 19.047 | 30.786 | -24.722 | -43.769 | 19.047 | 16.044 | 50.290 | 6-311++g(d) |
| $Ne^{2+}$ | 45.837 | 33.724 | 39.781 | 12.113 | 65.323 | -33.724 | -45.837 | 12.113 | 46.947 | 86.728 | 6-31+g(d) |
| | 45.766 | 33.711 | 39.738 | 12.056 | 65.494 | -33.711 | -45.766 | 12.056 | 47.132 | 86.871 | 6-311+g(d) |
| | 45.766 | 33.711 | 39.738 | 12.056 | 65.494 | -33.711 | -45.766 | 12.056 | 47.132 | 86.871 | 6-311++g(d) |

(d)

| Ions | I (eV) | E (eV) | χ (eV) | η (eV) | ω (eV) | μ+ (eV) | μ- (eV) | η (eV) | ω+ (eV) | ω- (eV) | Basis sets |
|---|---|---|---|---|---|---|---|---|---|---|---|
| $Li^-$ | -0.013 | -1.279 | -0.646 | 1.266 | 0.165 | 1.279 | 0.013 | 1.266 | 0.646 | 0.000 | 6-31+g(d) |
| | -0.016 | -1.241 | -0.628 | 1.224 | 0.161 | 1.241 | 0.016 | 1.225 | 0.628 | 0.000 | 6-311+g(d) |
| | -0.016 | -1.241 | -0.628 | 1.224 | 0.161 | 1.241 | 0.016 | 1.225 | 0.628 | 0.000 | 6-311++g(d) |
| $Be^-$ | 0.866 | -0.239 | 0.314 | 1.105 | 0.045 | 0.239 | -0.866 | 1.105 | 0.026 | 0.339 | 6-31+g(d) |
| | 0.943 | -0.234 | 0.354 | 1.176 | 0.053 | 0.234 | -0.943 | 1.177 | 0.023 | 0.378 | 6-311+g(d) |
| | 0.943 | -0.234 | 0.354 | 1.176 | 0.022 | 0.234 | -0.943 | 1.177 | 0.023 | 0.378 | 6-311++g(d) |
| $B^-$ | 0.931 | 1.301 | 1.116 | -0.370 | -1.683 | -1.301 | -0.931 | -0.370 | -2.288 | -1.171 | 6-31+g(d) |
| | 1.019 | 1.358 | 1.188 | -0.338 | -2.088 | -1.358 | -1.019 | -0.339 | -2.725 | -1.536 | 6-311+g(d) |
| | 1.019 | 1.358 | 1.188 | -0.338 | -2.088 | -1.358 | -1.019 | -0.339 | -2.725 | -1.536 | 6-311++g(d) |
| $C^-$ | 3.387 | 0.153 | 1.770 | 3.234 | 0.484 | -0.153 | -3.387 | 3.234 | 0.004 | 1.773 | 6-31+g(d) |
| | 3.440 | 0.169 | 1.805 | 3.271 | 0.497 | -0.169 | -3.440 | 3.271 | 0.004 | 1.809 | 6-311+g(d) |
| | 3.440 | 0.169 | 1.805 | 3.271 | 0.497 | -0.169 | -3.440 | 3.271 | 0.004 | 1.809 | 6-311++g(d) |
| $O^-$ | 6.154 | -0.200 | 2.977 | 6.354 | 0.697 | 0.200 | -6.154 | 6.354 | 0.003 | 2.980 | 6-31+g(d) |
| | 6.136 | -0.206 | 2.965 | 6.342 | 0.693 | 0.206 | -6.136 | 6.342 | 0.003 | 2.968 | 6-311+g(d) |
| | 6.136 | -0.206 | 2.965 | 6.342 | 0.693 | 0.206 | -6.136 | 6.342 | 0.003 | 2.968 | 6-311++g(d) |
| $F^-$ | 4.708 | -4.351 | 0.178 | 9.058 | 0.002 | 4.351 | -4.708 | 9.059 | 1.045 | 1.223 | 6-31+g(d) |
| | 4.684 | -4.459 | 0.112 | 9.143 | 0.001 | 4.459 | -4.684 | 9.143 | 1.087 | 1.199 | 6-311+g(d) |
| | 4.684 | -4.460 | 0.112 | 9.143 | 0.001 | 4.460 | -4.684 | 9.144 | 1.087 | 1.199 | 6-311++g(d) |
| $Ne^-$ | -1.359 | -5.847 | -5.104 | 1.488 | 8.750 | 5.847 | 1.359 | 1.488 | 11.488 | 6.348 | 6-31+g(d) |
| | -4.068 | -5.452 | -4.760 | 1.384 | 8.184 | 5.452 | 4.068 | 1.384 | 10.736 | 5.977 | 6-311+g(d) |
| | -4.068 | -5.452 | -4.760 | 1.384 | 8.184 | 5.452 | 4.068 | 1.384 | 10.736 | 5.977 | 6-311++g(d) |

(e)

| Ions | I (eV) | E (eV) | χ (eV) | η (eV) | ω (eV) | μ+ (eV) | μ- (eV) | η (eV) | ω+ (eV) | ω- (eV) | Basis sets |
|---|---|---|---|---|---|---|---|---|---|---|---|
| $Li^{2-}$ | -1.279 | -2.393 | -1.836 | 1.114 | 1.511 | 2.393 | 1.279 | 1.114 | 2.569 | 0.734 | 6-31+g(d) |
| | -1.241 | -2.371 | -1.806 | 1.131 | 1.442 | 2.371 | 1.241 | 1.130 | 2.487 | 0.681 | 6-311+g(d) |
| | -1.241 | -2.371 | -1.806 | 1.131 | 1.442 | 2.371 | 1.241 | 1.130 | 2.487 | 0.681 | 6-311++g(d) |
| $Be^{2-}$ | -0.239 | -0.103 | -0.171 | -0.136 | -0.107 | 0.103 | 0.239 | -0.136 | -0.039 | -0.210 | 6-31+g(d) |
| | -0.234 | -0.106 | -0.170 | -0.128 | -0.112 | 0.106 | 0.234 | -0.128 | -0.043 | -0.213 | 6-311+g(d) |
| | -0.234 | -0.106 | -0.170 | -0.128 | -0.112 | 0.106 | 0.234 | -0.128 | -0.043 | -0.213 | 6-311++g(d) |
| $B^{2-}$ | 1.301 | -0.040 | 0.631 | 1.341 | 0.148 | 0.040 | -1.301 | 1.341 | 0.001 | 0.631 | 6-31+g(d) |
| | 1.358 | 0.030 | 0.694 | 1.327 | 0.181 | -0.030 | -1.358 | 1.328 | 0.000 | 0.694 | 6-311+g(d) |
| | 1.358 | 0.030 | 0.694 | 1.327 | 0.181 | -0.030 | -1.358 | 1.328 | 0.000 | 0.694 | 6-311++g(d) |
| $C^{2-}$ | 0.153 | 0.871 | 0.512 | -0.719 | -0.182 | -0.871 | -0.153 | -0.718 | -0.528 | -0.016 | 6-31+g(d) |
| | 0.169 | 0.881 | 0.525 | -0.712 | -0.194 | -0.881 | -0.169 | -0.712 | -0.545 | -0.020 | 6-311+g(d) |
| | 0.169 | 0.881 | 0.525 | -0.712 | -0.194 | -0.881 | -0.169 | -0.712 | -0.545 | -0.020 | 6-311++g(d) |
| $O^{2-}$ | -0.200 | -2.451 | -1.326 | 2.251 | 0.390 | 2.451 | 0.200 | 2.251 | 1.335 | 0.009 | 6-31+g(d) |
| | -0.206 | -2.607 | -1.407 | 2.401 | 0.412 | 2.607 | 0.206 | 2.401 | 1.415 | 0.009 | 6-311+g(d) |
| | -0.206 | -2.607 | -1.406 | 2.401 | 0.412 | 2.607 | 0.206 | 2.401 | 1.415 | 0.009 | 6-311++g(d) |
| $F^{2-}$ | -4.351 | -5.134 | -4.742 | 0.783 | 14.366 | 5.134 | 4.351 | 0.783 | 16.835 | 12.093 | 6-31+g(d) |
| | -4.459 | -5.316 | -4.888 | 0.857 | 13.939 | 5.316 | 4.459 | 0.857 | 16.489 | 11.602 | 6-311+g(d) |
| | -4.459 | -5.316 | -4.888 | 0.857 | 13.939 | 5.316 | 4.459 | 0.857 | 16.489 | 11.602 | 6-311++g(d) |
| $Ne^{2-}$ | -5.488 | -6.485 | -6.166 | 0.638 | 29.811 | 6.485 | 5.488 | 0.637 | 32.974 | 26.807 | 6-31+g(d) |
| | -5.452 | -5.976 | -5.714 | 0.524 | 31.158 | 5.976 | 5.452 | 0.524 | 34.080 | 28.366 | 6-311+g(d) |
| | -5.452 | -5.976 | -5.714 | 0.524 | 31.158 | 5.976 | 5.452 | 0.524 | 34.080 | 28.366 | 6-311++g(d) |

**Table 8**: Ionization potential(I) , Electron affinity(E), Electronegativity(χ), Chemical hardness(η) , Electrophilicity(ω) and the values of ($\mu^+$, $\mu^-$, $\omega^+$, $\omega^-$ ) for Electroaccepting and Electrodonating processes of atoms and ions by using B3LYP method in aqueous phase

(a)

| Atoms | I (eV) | E (eV) | χ (eV) | η (eV) | ω (eV) | $\mu^+$ (eV) | $\mu^-$ (eV) | η (eV) | $\omega^+$ (eV) | $\omega^-$ (eV) | Basis sets |
|---|---|---|---|---|---|---|---|---|---|---|---|
| Li | 1.130 | 0.883 | 1.007 | 0.247 | 2.053 | -0.883 | -1.130 | 0.247 | 1.580 | 2.587 | 6-31+g(d) |
| | 1.137 | 0.870 | 1.004 | 0.267 | 1.888 | -0.870 | -1.137 | 0.267 | 1.419 | 2.423 | 6-311+g(d) |
| | 1.137 | 0.870 | 1.004 | 0.267 | 1.888 | -0.870 | -1.137 | 0.267 | 1.419 | 2.423 | 6-311++g(d) |
| Be | 4.624 | 1.423 | 3.024 | 3.202 | 1.428 | -1.423 | -4.624 | 3.201 | 0.316 | 3.340 | 6-31+g(d) |
| | 4.642 | 1.468 | 3.055 | 3.174 | 1.471 | -1.468 | -4.642 | 3.174 | 0.340 | 3.395 | 6-311+g(d) |
| | 4.642 | 1.468 | 3.055 | 3.174 | 1.471 | -1.468 | -4.642 | 3.174 | 0.340 | 3.395 | 6-311++g(d) |
| B | 5.832 | 2.074 | 3.953 | 3.758 | 2.079 | -2.074 | -5.832 | 3.758 | 0.572 | 4.526 | 6-31+g(d) |
| | 5.881 | 2.165 | 4.023 | 3.716 | 2.178 | -2.165 | -5.881 | 3.716 | 0.631 | 4.654 | 6-311+g(d) |
| | 5.881 | 2.165 | 4.023 | 3.716 | 2.178 | -2.165 | -5.881 | 3.716 | 0.631 | 4.654 | 6-311++g(d) |
| C | 6.745 | 4.340 | 5.543 | 2.404 | 6.389 | -4.340 | -6.745 | 2.405 | 3.918 | 9.461 | 6-31+g(d) |
| | 6.791 | 4.378 | 5.585 | 2.412 | 6.465 | -4.378 | -6.791 | 2.413 | 3.974 | 9.558 | 6-311+g(d) |
| | 6.791 | 4.378 | 5.585 | 2.412 | 6.465 | -4.378 | -6.791 | 2.413 | 3.974 | 9.558 | 6-311++g(d) |
| N | 11.436 | 4.108 | 7.772 | 7.329 | 4.121 | -4.108 | -11.436 | 7.328 | 1.151 | 8.923 | 6-31+g(d) |
| | 11.405 | 4.110 | 7.758 | 7.295 | 4.125 | -4.110 | -11.405 | 7.295 | 1.158 | 8.916 | 6-311+g(d) |
| | 11.405 | 4.110 | 7.758 | 7.295 | 4.125 | -4.110 | -11.405 | 7.295 | 1.158 | 8.916 | 6-311++g(d) |
| O | 11.981 | 7.593 | 9.787 | 4.388 | 10.914 | -7.593 | -11.981 | 4.388 | 6.569 | 16.356 | 6-31+g(d) |
| | 11.957 | 7.581 | 9.769 | 4.376 | 10.903 | -7.581 | -11.957 | 4.376 | 6.566 | 16.334 | 6-311+g(d) |
| | 11.957 | 7.581 | 9.769 | 4.376 | 10.903 | -7.581 | -11.957 | 4.376 | 6.566 | 16.334 | 6-311++g(d) |
| F | 17.908 | 6.913 | 12.410 | 10.996 | 7.004 | -6.913 | -17.908 | 10.996 | 2.173 | 14.583 | 6-31+g(d) |
| | 17.899 | 6.096 | 11.998 | 11.803 | 6.098 | -6.096 | -17.899 | 11.803 | 1.574 | 13.572 | 6-311+g(d) |
| | 17.899 | 6.096 | 11.998 | 11.803 | 6.098 | -6.096 | -17.899 | 11.803 | 1.574 | 13.572 | 6-311++g(d) |
| Ne | 18.219 | -3.347 | 7.436 | 21.567 | 1.282 | 3.347 | -18.219 | 21.566 | 0.260 | 7.696 | 6-31+g(d) |
| | 18.180 | -3.105 | 7.538 | 21.285 | 1.335 | 3.105 | -18.180 | 21.285 | 0.226 | 7.764 | 6-311+g(d) |
| | 18.180 | -3.105 | 7.538 | 21.285 | 1.335 | 3.105 | -18.180 | 21.285 | 0.226 | 7.764 | 6-311++g(d) |

(b)

| Ions | I (eV) | E (eV) | χ (eV) | η (eV) | ω (eV) | $\mu^+$ (eV) | $\mu^-$ (eV) | η (eV) | $\omega^+$ (eV) | $\omega^-$ (eV) | Basis sets |
|---|---|---|---|---|---|---|---|---|---|---|---|
| $Li^+$ | 62.692 | 1.130 | 31.911 | 61.562 | 8.271 | -1.130 | -62.692 | 61.562 | 0.010 | 31.922 | 6-31+g(d) |
| | 61.675 | 1.137 | 31.406 | 60.538 | 8.146 | -1.137 | -61.675 | 60.538 | 0.011 | 31.417 | 6-311+g(d) |
| | 61.675 | 1.137 | 31.406 | 60.538 | 8.146 | -1.137 | -61.675 | 60.538 | 0.011 | 31.417 | 6-311++g(d) |
| $Be^+$ | 6.036 | 4.624 | 5.331 | 1.412 | 10.064 | -4.624 | -6.036 | 1.412 | 7.576 | 10.906 | 6-31+g(d) |
| | 6.019 | 4.642 | 5.331 | 1.377 | 10.317 | -4.642 | -6.019 | 1.377 | 7.824 | 13.155 | 6-311+g(d) |
| | 6.019 | 4.642 | 5.331 | 1.377 | 10.317 | -4.642 | -6.019 | 1.377 | 7.824 | 13.155 | 6-311++g(d) |
| $B^+$ | 16.143 | 5.832 | 10.987 | 10.311 | 5.855 | -5.832 | -16.143 | 10.311 | 1.649 | 12.637 | 6-31+g(d) |
| | 16.116 | 5.881 | 10.999 | 10.237 | 5.912 | -5.881 | -16.116 | 10.237 | 1.689 | 12.689 | 6-311+g(d) |
| | 16.118 | 5.881 | 10.999 | 10.237 | 5.910 | -5.881 | -16.118 | 10.237 | 1.689 | 12.689 | 6-311++g(d) |
| $C^+$ | 15.885 | 6.745 | 11.315 | 9.140 | 7.004 | -6.745 | -15.885 | 9.140 | 2.488 | 13.804 | 6-31+g(d) |
| | 15.894 | 6.791 | 11.342 | 9.103 | 7.066 | -6.791 | -15.894 | 9.103 | 2.533 | 13.875 | 6-311+g(d) |
| | 15.894 | 6.791 | 11.342 | 9.103 | 7.066 | -6.791 | -15.894 | 9.103 | 2.533 | 13.875 | 6-311++g(d) |
| $N^+$ | 17.780 | 11.436 | 14.608 | 6.344 | 16.820 | -11.436 | -17.780 | 6.344 | 10.309 | 24.917 | 6-31+g(d) |
| | 17.717 | 11.405 | 14.561 | 6.312 | 16.796 | -11.405 | -17.717 | 6.312 | 10.305 | 24.866 | 6-311+g(d) |
| | 17.717 | 11.405 | 14.561 | 6.312 | 16.796 | -11.405 | -17.717 | 6.312 | 10.305 | 24.866 | 6-311++g(d) |
| $O^+$ | 24.953 | 11.981 | 18.467 | 12.972 | 13.145 | -11.981 | -24.953 | 12.972 | 5.533 | 23.999 | 6-31+g(d) |
| | 24.880 | 11.957 | 18.418 | 12.923 | 13.126 | -11.957 | -24.880 | 12.923 | 5.532 | 23.950 | 6-311+g(d) |
| | 24.880 | 11.957 | 18.418 | 12.923 | 13.126 | -11.957 | -24.880 | 12.923 | 5.532 | 23.950 | 6-311++g(d) |
| $F^+$ | 26.522 | 17.908 | 22.215 | 8.614 | 28.646 | -17.908 | -26.522 | 8.614 | 18.616 | 40.831 | 6-31+g(d) |
| | 26.450 | 17.899 | 22.175 | 8.551 | 28.753 | -17.899 | -26.450 | 8.551 | 18.734 | 40.909 | 6-311+g(d) |
| | 26.450 | 17.899 | 22.175 | 8.551 | 28.753 | -17.899 | -26.450 | 8.551 | 18.734 | 40.909 | 6-311++g(d) |
| $Ne^+$ | 34.992 | 18.219 | 26.606 | 16.773 | 21.102 | -18.219 | -34.992 | 16.773 | 9.895 | 36.501 | 6-31+g(d) |
| | 34.939 | 18.180 | 26.560 | 8.379 | 21.046 | -18.180 | -34.939 | 16.759 | 9.861 | 36.421 | 6-311+g(d) |
| | 34.939 | 18.180 | 26.560 | 8.379 | 21.046 | -18.180 | -34.939 | 16.759 | 9.861 | 36.421 | 6-311++g(d) |

(c)

| Ions | I (eV) | E (eV) | χ (eV) | η (eV) | ω (eV) | $\mu^+$ (eV) | $\mu^-$ (eV) | η (eV) | $\omega^+$ (eV) | $\omega^-$ (eV) | Basis sets |
|---|---|---|---|---|---|---|---|---|---|---|---|
| $Li^{2+}$ | -62.421 | 62.692 | 0.136 | -125.113 | -0.000 | -62.692 | 62.421 | -125.113 | -15.707 | -15.571 | 6-31+g(d) |
| | -61.511 | 61.675 | 0.082 | -123.187 | -0.000 | -61.675 | 61.511 | -123.186 | -15.439 | -15.357 | 6-311+g(d) |
| | -61.511 | 61.675 | 0.082 | -123.187 | -0.000 | -61.675 | 61.511 | -123.186 | -15.439 | -15.357 | 6-311++g(d) |
| $Be^{2+}$ | 133.331 | 6.036 | 69.684 | 127.285 | 19.073 | -6.036 | -133.331 | 127.285 | 0.143 | 69.827 | 6-31+g(d) |
| | 132.869 | 6.019 | 69.444 | 126.849 | 19.009 | -6.019 | -132.869 | 126.850 | 0.143 | 69.587 | 6-311+g(d) |
| | 132.869 | 6.019 | 69.444 | 126.849 | 19.009 | -6.019 | -132.869 | 126.850 | 0.143 | 69.587 | 6-311++g(d) |
| $B^{2+}$ | 24.048 | 16.143 | 20.096 | 7.906 | 25.541 | -16.143 | -24.048 | 7.905 | 16.481 | 36.577 | 6-31+g(d) |
| | 24.071 | 16.118 | 20.095 | 7.953 | 25.385 | -16.118 | -24.071 | 7.953 | 16.332 | 36.427 | 6-311+g(d) |
| | 24.071 | 16.118 | 20.095 | 7.953 | 25.385 | -16.118 | -24.071 | 7.953 | 16.332 | 36.427 | 6-311++g(d) |
| $C^{2+}$ | 31.973 | 15.885 | 23.929 | 16.088 | 17.796 | -15.885 | -31.973 | 16.088 | 7.842 | 31.772 | 6-31+g(d) |
| | 32.007 | 15.894 | 23.951 | 16.114 | 17.799 | -15.894 | -32.007 | 16.113 | 7.838 | 31.789 | 6-311+g(d) |
| | 32.007 | 15.894 | 23.951 | 16.114 | 17.799 | -15.894 | -32.007 | 16.113 | 7.838 | 31.789 | 6-311++g(d) |
| $N^{2+}$ | 32.231 | 17.780 | 25.006 | 14.451 | 21.634 | -17.780 | -32.231 | 14.451 | 10.938 | 35.944 | 6-31+g(d) |
| | 32.299 | 17.717 | 25.008 | 14.582 | 21.444 | -17.717 | -32.299 | 14.582 | 10.763 | 35.771 | 6-311+g(d) |
| | 32.299 | 17.717 | 25.008 | 14.581 | 21.444 | -17.717 | -32.299 | 14.582 | 10.763 | 35.771 | 6-311++g(d) |
| $O^{2+}$ | 35.255 | 24.953 | 30.104 | 10.302 | 43.983 | -24.953 | -35.255 | 10.302 | 30.219 | 60.323 | 6-31+g(d) |
| | 35.156 | 24.880 | 30.018 | 10.276 | 43.842 | -24.880 | -35.156 | 10.276 | 30.117 | 60.135 | 6-311+g(d) |
| | 35.156 | 24.880 | 30.018 | 10.276 | 43.842 | -24.880 | -35.156 | 10.276 | 30.117 | 60.135 | 6-311++g(d) |
| $F^{2+}$ | 44.948 | 26.522 | 35.735 | 18.426 | 34.652 | -26.522 | -44.948 | 18.426 | 19.088 | 54.823 | 6-31+g(d) |
| | 44.868 | 26.450 | 35.659 | 18.418 | 34.520 | -26.450 | -44.868 | 18.418 | 18.993 | 54.652 | 6-311+g(d) |
| | 44.868 | 26.450 | 35.659 | 18.418 | 34.520 | -26.450 | -44.868 | 18.418 | 18.993 | 54.652 | 6-311++g(d) |
| $Ne^{2+}$ | 47.597 | 34.992 | 41.295 | 12.605 | 67.644 | -34.992 | -47.597 | 12.605 | 48.572 | 89.867 | 6-31+g(d) |
| | 47.507 | 34.939 | 41.223 | 12.567 | 67.609 | -34.939 | -47.507 | 12.568 | 48.568 | 89.792 | 6-311+g(d) |
| | 47.507 | 34.939 | 41.223 | 12.567 | 67.609 | -34.939 | -47.507 | 12.568 | 48.568 | 89.792 | 6-311++g(d) |

(d)

| Ions | I (eV) | E (eV) | χ (eV) | η (eV) | ω (eV) | $\mu^+$ (eV) | $\mu^-$ (eV) | η (eV) | $\omega^+$ (eV) | $\omega^-$ (eV) | Basis sets |
|---|---|---|---|---|---|---|---|---|---|---|---|
| $Li^-$ | 0.883 | -0.758 | 0.062 | 1.641 | 0.001 | 0.758 | -0.883 | 1.641 | 0.175 | 0.237 | 6-31+g(d) |
| | 0.870 | -0.709 | 0.081 | 1.580 | 0.002 | 0.709 | -0.870 | 1.579 | 0.159 | 0.240 | 6-311+g(d) |
| | 0.870 | -0.709 | 0.081 | 1.580 | 0.002 | 0.709 | -0.870 | 1.579 | 0.159 | 0.240 | 6-311++g(d) |
| $Be^-$ | 1.423 | 0.653 | 1.038 | 0.699 | 0.699 | -0.653 | -1.423 | 0.770 | 0.276 | 1.314 | 6-31+g(d) |
| | 1.468 | 0.655 | 1.061 | 0.814 | 0.692 | -0.655 | -1.468 | 0.813 | 0.263 | 1.325 | 6-311+g(d) |
| | 1.468 | 0.655 | 1.061 | 0.814 | 0.692 | -0.655 | -1.468 | 0.813 | 0.263 | 1.325 | 6-311++g(d) |
| $B^-$ | 2.074 | 1.904 | 1.989 | 0.169 | 11.669 | -1.904 | -2.074 | 0.170 | 10.696 | 12.685 | 6-31+g(d) |
| | 2.165 | 1.965 | 2.065 | 0.200 | 10.644 | -1.965 | -2.165 | 0.200 | 9.636 | 11.701 | 6-311+g(d) |
| | 2.165 | 1.965 | 2.065 | 0.200 | 10.644 | -1.965 | -2.165 | 0.200 | 9.636 | 11.701 | 6-311++g(d) |
| $C^-$ | 4.341 | 1.621 | 2.981 | 2.719 | 1.633 | -1.621 | -4.341 | 2.720 | 0.483 | 3.454 | 6-31+g(d) |
| | 4.378 | 1.630 | 3.004 | 2.749 | 1.642 | -1.630 | -4.378 | 2.748 | 0.483 | 3.487 | 6-311+g(d) |
| | 4.378 | 1.630 | 3.004 | 2.749 | 1.642 | -1.630 | -4.378 | 2.748 | 0.483 | 3.487 | 6-311++g(d) |
| $N^-$ | 4.108 | 3.288 | 3.698 | 0.819 | 8.348 | -3.288 | -4.108 | 0.819 | 6.601 | 10.299 | 6-31+g(d) |
| | 4.110 | 3.264 | 3.687 | 0.846 | 8.029 | -3.264 | -4.110 | 0.846 | 6.292 | 9.978 | 6-311+g(d) |
| | 4.110 | 3.264 | 3.687 | 0.846 | 8.029 | -3.264 | -4.110 | 0.846 | 6.292 | 9.978 | 6-311++g(d) |
| $O^-$ | 6.913 | 2.082 | 4.837 | 5.511 | 2.123 | -2.082 | -6.913 | 5.511 | 0.393 | 5.231 | 6-31+g(d) |
| | 7.581 | 2.063 | 4.822 | 5.518 | 2.107 | -2.063 | -7.581 | 5.518 | 0.386 | 5.208 | 6-311+g(d) |
| | 7.582 | 2.063 | 4.822 | 5.518 | 2.107 | -2.063 | -7.582 | 5.519 | 0.386 | 5.208 | 6-311++g(d) |
| $F^-$ | 6.913 | -3.306 | 1.803 | 10.218 | 0.159 | 3.306 | -6.913 | 10.219 | 0.535 | 2.338 | 6-31+g(d) |
| | 6.898 | -3.439 | 1.729 | 8.734 | 0.171 | 3.439 | -6.898 | 8.737 | 0.398 | 2.128 | 6-311+g(d) |
| | 6.096 | -2.637 | 1.729 | 8.734 | 0.171 | 2.637 | -6.096 | 8.733 | 0.398 | 2.128 | 6-311++g(d) |
| $Ne^-$ | -3.347 | -4.377 | -3.863 | 1.030 | 7.242 | 4.377 | 3.347 | 1.030 | 9.302 | 5.44. | 6-31+g(d) |
| | -3.103 | -4.031 | -3.568 | 0.926 | 6.873 | 4.031 | 3.103 | 0.928 | 8.773 | 5.205 | 6-311+g(d) |
| | -3.105 | -4.031 | -3.568 | 0.926 | 6.873 | 4.031 | 3.105 | 0.926 | 8.773 | 5.205 | 6-311++g(d) |

(e)

| **Ions** | **I (eV)** | **E (eV)** | **χ (eV)** | **η (eV)** | **ω (eV)** | **$\mu^+$ (eV)** | **$\mu^-$ (eV)** | **η (eV)** | **$\omega^+$ (eV)** | **$\omega^-$ (eV)** | **Basis sets** |
|---|---|---|---|---|---|---|---|---|---|---|---|
| $Li^{2-}$ | -0.758 | -1.527 | -1.143 | 0.768 | 0.849 | 1.527 | 0.758 | 0.769 | 1.517 | 0.374 | 6-31+g(d) |
| | -0.709 | -1.462 | -1.086 | 0.753 | 0.783 | 1.462 | 0.709 | 0.753 | 1.420 | 0.334 | 6-311+g(d) |
| | -0.709 | -1.462 | -1.086 | 0.753 | 0.783 | 1.462 | 0.709 | 0.753 | 1.420 | 0.334 | 6-311++g(d) |
| $Be^{2-}$ | 0.653 | 0.439 | 0.546 | 0.214 | 0.697 | -0.439 | -0.653 | 0.214 | 0.451 | 0.997 | 6-31+g(d) |
| | 0.655 | 0.416 | 0.535 | 0.238 | 0.601 | -0.416 | -0.655 | 0.239 | 0.363 | 0.898 | 6-311+g(d) |
| | 0.655 | 0.416 | 0.535 | 0.238 | 0.601 | -0.416 | -0.655 | 0.239 | 0.363 | 0.898 | 6-311++g(d) |
| $B^{2-}$ | 1.905 | 0.978 | 1.441 | 0.936 | 1.121 | -0.978 | -1.905 | 0.927 | 0.516 | 1.958 | 6-31+g(d) |
| | 1.965 | 1.040 | 1.502 | 0.925 | 1.220 | -1.040 | -1.965 | 0.925 | 0.585 | 2.087 | 6-311+g(d) |
| | 1.965 | 1.040 | 1.502 | 0.925 | 1.220 | -1.040 | -1.965 | 0.925 | 0.585 | 2.087 | 6-311++g(d) |
| $C^{2-}$ | 1.621 | 1.836 | 1.728 | -0.215 | -6.995 | -1.836 | -1.621 | -0.215 | -7.843 | -6.115 | 6-31+g(d) |
| | 1.630 | 1.831 | 1.731 | -0.202 | -7.426 | -1.831 | -1.630 | -0.201 | -8.316 | -6.586 | 6-311+g(d) |
| | 1.630 | 1.831 | 1.731 | -0.202 | -7.426 | -1.831 | -1.630 | -0.201 | -8.316 | -6.586 | 6-311++g(d) |
| $N^{2-}$ | 3.288 | 0.959 | 2.124 | 2.329 | 0.968 | -0.959 | -3.288 | 2.329 | 0.197 | 2.321 | 6-31+g(d) |
| | 3.263 | 0.946 | 2.105 | 2.317 | 0.956 | -0.946 | -3.263 | 2.317 | 0.193 | 2.298 | 6-311+g(d) |
| | 3.263 | 0.946 | 2.105 | 2.317 | 0.956 | -0.946 | -3.263 | 2.317 | 0.193 | 2.298 | 6-311++g(d) |
| $O^{2-}$ | 2.082 | -2.064 | 0.009 | 4.146 | 0.000 | 2.064 | -2.082 | 4.146 | 0.614 | 0.523 | 6-31+g(d) |
| | 2.063 | -2.286 | -0.111 | 4.349 | 0.001 | 2.286 | -2.063 | 4.349 | 0.601 | 0.489 | 6-311+g(d) |
| | 2.063 | -2.286 | -0.111 | 4.349 | 0.001 | 2.286 | -2.063 | 4.349 | 0.601 | 0.489 | 6-311++g(d) |
| $F^{2-}$ | -3.306 | -3.664 | -3.483 | 0.355 | 17.105 | 3.664 | 3.306 | 0.358 | 18.891 | 15.408 | 6-31+g(d) |
| | -3.439 | -3.865 | -3.653 | 0.425 | 15.677 | 3.865 | 3.439 | 0.425 | 17.557 | 13.905 | 6-311+g(d) |
| | -2.637 | -3.865 | -3.251 | 1.227 | 4.306 | 3.865 | 2.637 | 1.228 | 6.085 | 2.834 | 6-311++g(d) |
| $Ne^{2-}$ | -4.377 | -5.974 | -5.086 | 1.416 | 9.131 | 5.974 | 4.377 | 1.417 | 11.851 | 6.765 | 6-31+g(d) |
| | -4.031 | -5.274 | -4.653 | 1.243 | 8.705 | 5.274 | 4.031 | 1.243 | 11.187 | 6.534 | 6-311+g(d) |
| | -4.031 | -5.274 | -4.653 | 1.243 | 8.705 | 5.274 | 4.031 | 1.243 | 11.187 | 6.534 | 6-311++g(d) |

**Table – 9**: Comparison of the calculated values with the experimental values*

| Atoms/Ions | IP[a] | EA[a] | $\chi$[a] | $\eta$[a] |
|---|---|---|---|---|
| Li | 5.62(5.39) | 0.56(0.62) | 3.09(3.00) | 2.52(2.38) |
| Be | 9.12(9.30) | -0.23(0.40) | 4.45(4.90) | 4.67(4.50) |
| B | 8.73(8.30) | -0.35(0.28) | 4.19(4.29) | 4.06(4.01) |
| C | 9.77(11.26) | 1.64(1.27) | 5.76(6.27) | 5.46(5.00) |
| N | 14.60(14.53) | 1.06(0.07) | 7.83(7.30) | 6.72(7.23) |
| O | 15.29(13.62) | 4.36(1.46) | 9.82(7.54) | 5.97(6.08) |
| F | 21.40(17.42) | 3.48(3.40) | 12.44(10.41) | 8.96(7.01) |
| $Li^+$ | 76.05(75.64) | 5.62(5.39) | 40.93(40.52) | 35.22(35.12) |
| | **IP[b]** | **EA[b]** | **$\chi$[b]** | **$\eta$[b]** |
| Li | 5.617(5.39) | 0.56(0.62) | 3.09(3.01) | 2.52(2.39) |
| Be | 9.12(9.3) | -0.23(0.4) | 4.45(4.9) | 4.67(4.5) |
| B | 8.73(8.3) | -0.35(0.28) | 4.19(4.29) | 4.06(4.01) |
| C | 9.77(11.26) | 1.64(1.27) | 5.76(6.27) | 5.46(5.00) |
| N | 14.60(14.53) | 1.06(0.07) | 7.83(7.30) | 6.72(7.23) |
| O | 15.29(13.62) | 4.36(1.46) | 9.82(7.54) | 5.97(6.08) |
| F | 21.40(17.42) | 3.48(3.40) | 12.44(10.41) | 8.96(7.01) |
| $Li^+$ | 76.05(75.64) | 5.62(5.39) | 40.83(40.52) | 35.22(35.12) |
| $Be^{2+}$ | 154.26(153.89) | 18.59(18.21) | 86.42(86.05) | 67.83(67.84) |

* All the values are in eV. Experimental values (within parentheses) are taken from the references. 18 and 20. Experimental values have been taken from the ref. 18. Experimental values have been taken from the ref. 20. Calculated values are taken from the gas phase calculation at the B3LYP/6-311+G(d) level of theory.

**Table 10**: Ionization potential(I) , Electron affinity(A), Electronegativity($\chi$), Chemical hardness($\eta$) , Electrophilicity($\omega$) and the values of ($\mu^+$, $\mu^-$ ,$\omega^+$ ,$\omega^-$ ) for Electroaccepting and Electrodonating processes of the dianions in the presence of counterion($Z^+$) calculated at the HF/6-311+G(d) level of theory.

| Ions | I (eV) | A (eV) | $\chi$ (eV) | $\eta$ (eV) | $\omega$ (eV) | $\mu^+$ (eV) | $\mu^-$ (eV) | $\eta$ (eV) | $\omega^+$ (eV) | $\omega^-$ (eV) |
|---|---|---|---|---|---|---|---|---|---|---|
| $Li^{2-}(Z^+)_2$ | 5.338 | -0.124 | 2.731 | 5.462 | 0.683 | 0.124 | -5.339 | 5.462 | 0.002 | 2.609 |
| $Be^{2-}(Z^+)_2$ | 8.047 | -0.703 | 4.375 | 8.751 | 1.094 | 0.703 | -8.047 | 8.751 | 0.028 | 3.700 |
| $B^{2-}(Z^+)_2$ | 8.022 | -1.574 | 4.798 | 9.596 | 1.1995 | 1.574 | -8.022 | 9.596 | 0.129 | 3.353 |
| $C^{2-}(Z^+)_2$ | 8.378 | 0.694 | 3.842 | 7.684 | 0.961 | -0.694 | -8.378 | 7.684 | 0.031 | 4.568 |
| $N^{2-}(Z^+)_2$ | 13.586 | -0.729 | 7.157 | 14.315 | 1.789 | 0.729 | -13.586 | 14.315 | 0.018 | 6.447 |
| $O^{2-}(Z^+)_2$ | 13.576 | 2.874 | 5.351 | 10.703 | 1.338 | -2.874 | -13.576 | 10.703 | 0.386 | 8.611 |
| $F^{2-}(Z^+)_2$ | 20.092 | 1.245 | 9.423 | 18.846 | 2.356 | -1.245 | -20.092 | 18.846 | 0.041 | 10.710 |
| $Ne^{2-}(Z^+)_2$ | 19.794 | -7.553 | 13.673 | 27.346 | 3.418 | 7.553 | -19.794 | 27.346 | 1.043 | 7.164 |

**Table 11**: Ionization potential(I) , Electron affinity(A), Electronegativity(χ), Chemical hardness(η) , Electrophilicity(ω) and the values of ($\mu^+$, $\mu^-$ ,$\omega^+$ ,$\omega^-$ ) for Electroaccepting and Electrodonating processes of the dianions in the presence of counterion($Z^+$) calculated at the MP2/6-311+G(d) level of theory.

| **Ions** | **I (eV)** | **A (eV)** | **χ (eV)** | **η (eV)** | **ω (eV)** | **$\mu^+$ (eV)** | **$\mu^-$ (eV)** | **η (eV)** | **$\omega^+$ (eV)** | **$\omega^-$ (eV)** |
|---|---|---|---|---|---|---|---|---|---|---|
| $Li^{2-}(Z^+)_2$ | 5.338 | -0.124 | 2.731 | 5.462 | 0.683 | 0.124 | -5.338 | 5.462 | 0.001 | 2.609 |
| $Be^{2-}(Z^+)_2$ | 8.047 | -0.703 | 4.375 | 8.751 | 1.094 | 0.703 | -8.047 | 8.751 | 0.028 | 3.700 |
| $B^{2-}(Z^+)_2$ | 8.022 | -1.574 | 4.798 | 9.596 | 1.199 | 1.574 | -8.022 | 9.596 | 0.129 | 3.353 |
| $C^{2-}(Z^+)_2$ | 8.378 | 0.694 | 3.842 | 7.684 | 0.961 | -0.694 | -8.378 | 7.684 | 0.031 | 4.568 |
| $N^{2-}(Z^+)_2$ | 13.586 | -0.729 | 7.157 | 14.315 | 1.789 | 0.729 | -13.586 | 14.315 | 0.018 | 6.447 |
| $O^{2-}(Z^+)_2$ | 13.576 | 2.874 | 5.351 | 10.702 | 1.338 | -2.874 | -13.576 | 10.702 | 0.386 | 8.611 |
| $F^{2-}(Z^+)_2$ | 20.092 | 1.245 | 9.423 | 18.846 | 2.356 | -1.245 | -20.092 | 18.846 | 0.041 | 10.710 |
| $Ne^{2-}(Z^+)_2$ | 19.794 | -7.553 | 13.673 | 27.346 | 3.418 | 7.553 | -19.794 | 27.346 | 1.043 | 7.164 |

**Table 12**: Ionization potential(I) , Electron affinity(A), Electronegativity(χ), Chemical hardness(η) , Electrophilicity(ω) and the values of ($\mu^+$, $\mu^-$ ,$\omega^+$ ,$\omega^-$ ) for Electroaccepting and Electrodonating processes of the dianions in the presence of counterion($Z^+$) calculated at the B3LYP/6-311+G(d) level of theory.

| **Ions** | **I (eV)** | **A (eV)** | **χ (eV)** | **η (eV)** | **ω (eV)** | **$\mu^+$ (eV)** | **$\mu^-$ (eV)** | **η (eV)** | **$\omega^+$ (eV)** | **$\omega^-$ (eV)** |
|---|---|---|---|---|---|---|---|---|---|---|
| $Li^{2-}(Z^+)_2$ | 5.617 | 0.558 | 3.087 | 5.059 | 0.942 | -0.558 | -5.617 | 5.059 | 0.031 | 3.118 |
| $Be^{2-}(Z^+)_2$ | 9.117 | -0.228 | 4.445 | 9.345 | 1.057 | 0.228 | -9.117 | 9.345 | 0.003 | 4.447 |
| $B^{2-}(Z^+)_2$ | 8.728 | -0.350 | 4.189 | 9.078 | 0.966 | 0.350 | -8.728 | 9.079 | 0.007 | 4.196 |
| $C^{2-}(Z^+)_2$ | 9.771 | 1.642 | 5.706 | 8.128 | 2.003 | -1.642 | -9.771 | 8.128 | 0.166 | 5.872 |
| $N^{2-}(Z^+)_2$ | 14.603 | 1.058 | 7.831 | 13.544 | 2.264 | -1.058 | -14.603 | 13.544 | 0.041 | 7.872 |
| $O^{2-}(Z^+)_2$ | 15.293 | 4.357 | 9.825 | 10.936 | 4.413 | -4.357 | -15.293 | 10.936 | 0.868 | 10.693 |
| $F^{2-}(Z^+)_2$ | 21.397 | 3.486 | 12.442 | 17.911 | 4.321 | -3.486 | -21.397 | 17.911 | 0.339 | 12.781 |
| $Ne^{2-}(Z^+)_2$ | 21.802 | -6.584 | 7.609 | 28.386 | 1.020 | 6.584 | -21.802 | 28.386 | 0.764 | 8.373 |

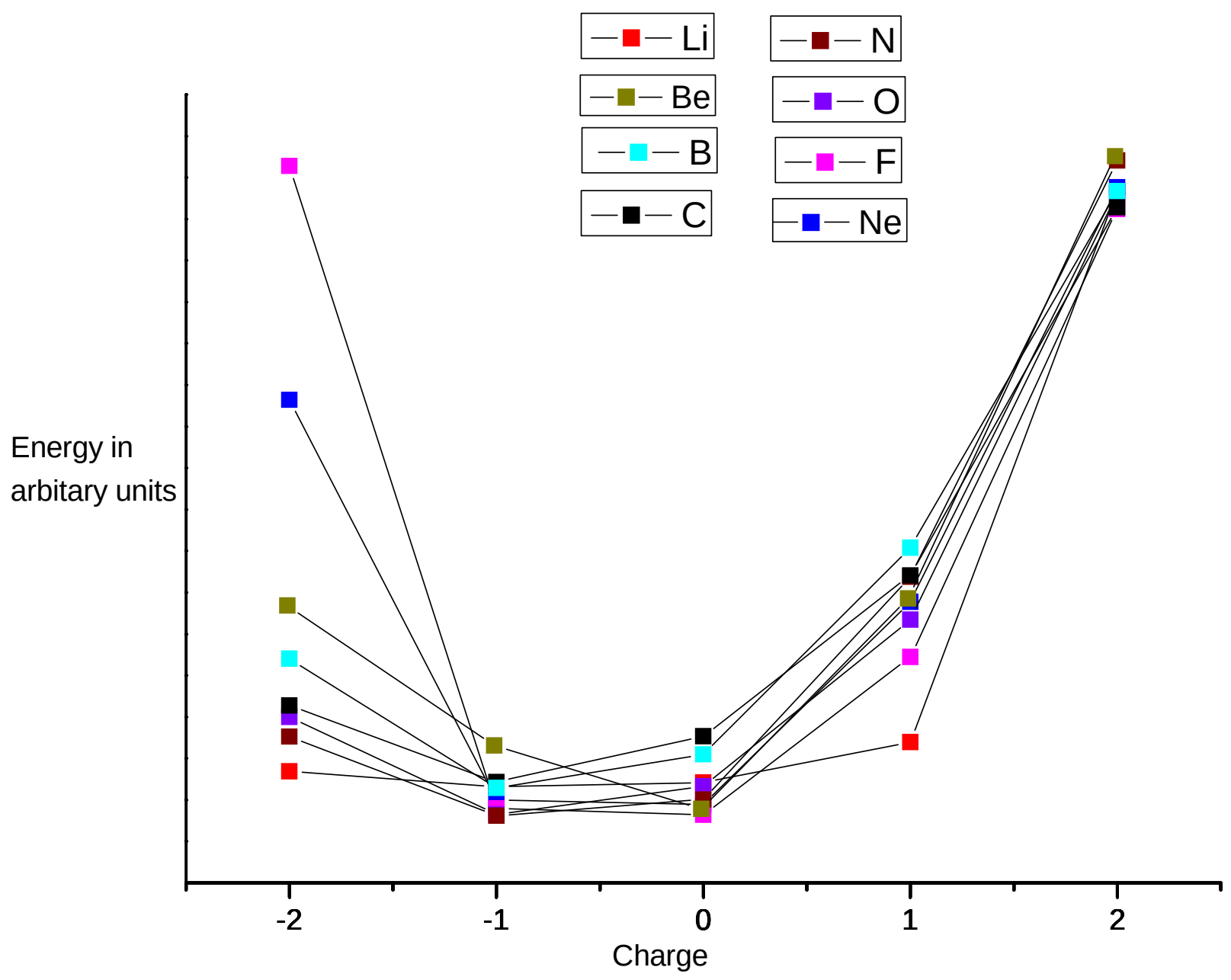

**Figure 1:** Changes in Energy values of Li through Ni with charges (-2 to +2)